\documentclass[natbib]{ar-1col}
\usepackage[numbers]{natbib}
\usepackage{amsmath}
\usepackage{amssymb}
\usepackage{amstext}
\usepackage{graphicx}
\usepackage{epstopdf}
\usepackage{epsfig}
\usepackage{url}
\usepackage{natbib}
\usepackage{subcaption} 
\usepackage{float} 
\usepackage{adjustbox}
\usepackage{lscape}
\usepackage{threeparttable}
\usepackage{tabulary}
\newcolumntype{K}[1]{>{\centering\arraybackslash}p{#1}}
\usepackage{cellspace}
\usepackage{longtable}

\graphicspath{{figures/}}

\begin{document}

\def\aj{AJ}%
\def\araa{ARA\&A}%
\def\apj{ApJ}%
\def\apjl{ApJ}%
\def\apjs{ApJS}%
\def\ao{Appl.~Opt.}%
\def\apss{Ap\&SS}%
\def\aap{A\&A}%
\def\aapr{A\&A~Rev.}%
\def\aaps{A\&AS}%
\def\azh{AZh}%
\def\baas{BAAS}%
\def\jrasc{JRASC}%
\def\memras{MmRAS}%
\def\mnras{MNRAS}%
\def\pra{Phys.~Rev.~A}%
\def\prb{Phys.~Rev.~B}%
\def\prc{Phys.~Rev.~C}%
\def\prd{Phys.~Rev.~D}%
\def\pre{Phys.~Rev.~E}%
\def\prl{Phys.~Rev.~Lett.}%
\def\pasp{PASP}%
\def\pasj{PASJ}%
\def\qjras{QJRAS}%
\def\skytel{S\&T}%
\def\solphys{Sol.~Phys.}%
\def\sovast{Soviet~Ast.}%
\def\ssr{Space~Sci.~Rev.}%
\def\zap{ZAp}%
\def\nat{Nature}%
\def\iaucirc{IAU~Circ.}%
\def\aplett{Astrophys.~Lett.}%
\def\apspr{Astrophys.~Space~Phys.~Res.}%
\def\bain{Bull.~Astron.~Inst.~Netherlands}%
\def\fcp{Fund.~Cosmic~Phys.}%
\def\gca{Geochim.~Cosmochim.~Acta}%
\def\grl{Geophys.~Res.~Lett.}%
\def\icarus{Icarus}%
\def\jcp{J.~Chem.~Phys.}%
\def\jgr{J.~Geophys.~Res.}%
\def\jqsrt{J.~Quant.~Spec.~Radiat.~Transf.}%
\def\memsai{Mem.~Soc.~Astron.~Italiana}%
\def\nphysa{Nucl.~Phys.~A}%
\def\physrep{Phys.~Rep.}%
\def\physscr{Phys.~Scr}%
\def\planss{Planet.~Space~Sci.}%
\def\procspie{Proc.~SPIE}%
\def\rmxaa{Rev.~Mex.~Astron.~Astrofis.}
\def\jcap{JCAP}
\let\astap=\aap
\let\apjlett=\apjl
\let\apjsupp=\apjs
\let\applopt=\ao

\markboth{N. Madhusudhan}{Exoplanetary Atmospheres}

\title{Exoplanetary Atmospheres: Key Insights, Challenges and Prospects}

\author{Nikku Madhusudhan$^1$ 
\affil{$^1$Institute of Astronomy, University of Cambridge, Cambridge, UK, CB3 0HA; email: nmadhu@ast.cam.ac.uk}}

\begin{abstract}
Exoplanetary science is on the verge of an unprecedented revolution. The thousands of exoplanets discovered over the past decade have most recently been supplemented by discoveries of potentially habitable planets around nearby low-mass stars. Currently, the field is rapidly progressing towards detailed spectroscopic observations to characterise the atmospheres of these planets. While various surveys from space and ground are expected to detect numerous more exoplanets orbiting nearby stars, the imminent launch of JWST along with large ground-based facilities are expected to revolutionise exoplanetary spectroscopy. Such observations, combined with detailed theoretical models and inverse methods, provide valuable insights into a wide range of physical processes and chemical compositions in exoplanetary atmospheres. Depending on the planetary properties, the knowledge of atmospheric compositions can also place important constraints on planetary formation and migration mechanisms, geophysical processes and, ultimately, biosignatures. In the present review, we will discuss the modern and future landscape of this frontier area of exoplanetary atmospheres. We will start with a brief review of the area, emphasising the key insights gained from different observational methods and theoretical studies. This is followed by an in-depth discussion of the state-of-the-art, challenges, and future prospects in three forefront branches of the area. 
\end{abstract}

\begin{keywords}
Extrasolar planets, exoplanetary atmospheres, planet formation, habitability, atmospheric composition
\end{keywords}
\maketitle

\tableofcontents

\section{Introduction} 
\label{sec:intro}

Planetary atmospheres serve as Rosetta Stones for planetary processes. Encoded in the spectrum of a planetary atmosphere is information about its various physical and chemical properties which in turn provide key insights into myriad atmospheric processes as well as the formation and evolutionary history of the planet. Over a century of spectroscopic observations of planets and moons in the solar system have revealed a vast treasury of information on their diversity in all these aspects. From the giant storms and NH$_3$ clouds on Jupiter to the dense sulphuric clouds on Venus, from H$_2$-rich giant planets to CO2-rich Venus and Mars, and then the unique Earth, the atmospheric diversity of the solar system is a sight to behold for the intrepid explorer. Yet, all the breathtaking diversity of the solar system arises from a surprisingly limited range of macroscopic planetary parameters from a cosmic context. The equilibrium temperatures of the solar system planets lie between 50 and 500 K, with only Venus and Mercury being above 300 K. The planetary sizes and masses fall in three broad ranges - the gas giants (8-11 R$_\oplus$, $\sim$100-320M$_\oplus$), the ice giants (4R$_\oplus$, $\sim$14-17M$_\oplus$), and the terrestrial planets ($\leq$ 1R$_\oplus$,$\leq$ 1M$_\oplus$). In contrast, the thousands of exoplanets known today span a range in bulk properties that would have been considered impossible two decades ago with temperatures ranging between 200 - 4000 K and radii and masses spanning continuously over a large range ($\sim$0.5-20 R$_\oplus$,$\sim$1-10$^4$ M$_\oplus$). It is only natural then that the atmospheres of these exoplanets may also be expected to be similarly diverse. As such, exoplanetary atmospheres serve as excellent laboratories to study planetary processes and formation mechanisms over the entire possible range of macroscopic properties - masses, radii, temperatures/irradiation, orbital architectures, and host stellar types.

\begin{figure}[t]
\includegraphics[width=1.0\linewidth]{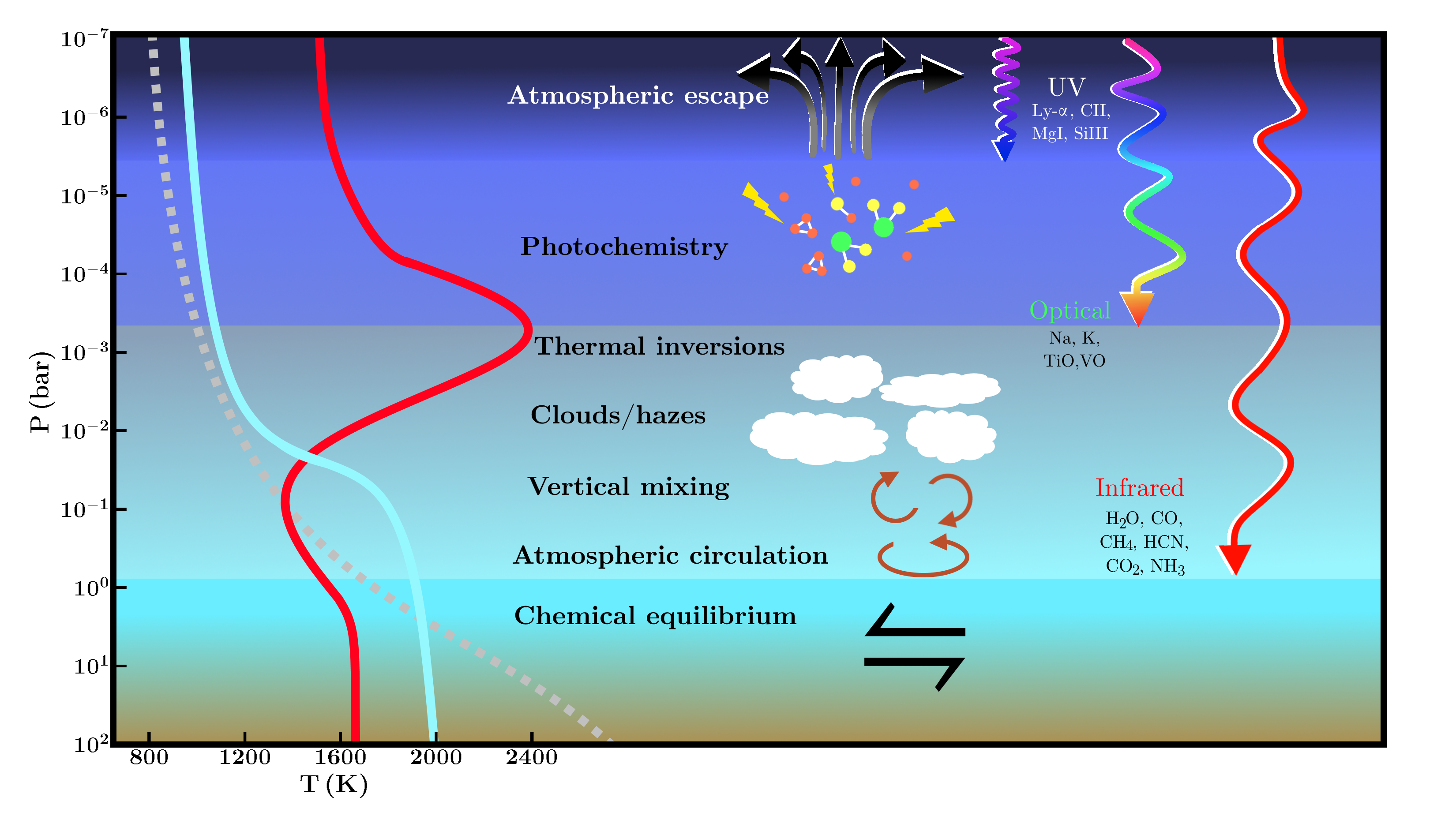}
\caption{Processes active in exoplanetary atmospheres and how they are probed by different parts of the electromagnetic spectrum. These processes occur in different regions of the atmosphere and are labelled at the relevant location. On the right, the penetration depths of UV, optical and infrared light are shown, indicating which processes can be probed by observations in each wavelength range. The chemical species whose signatures can be detected in each wavelength range are also indicated. On the left are shown three types of temperature profile which can arise as a result of atmospheric processes: the profile of a highly-irradiated planet with a thermal inversion (red), that of an irradiated planet without a thermal inversion (cyan), and the temperature profile of a poorly irradiated planet (grey, dashed).}
\label{fig:atm_schematic}
\end{figure}

The information obtained about an exoplanetary atmosphere depends on the nature of observations. Figure~\ref{fig:atm_schematic} shows a schematic of atmospheric processes that can be observed in exoplanets with different spectral ranges probing different regions, and hence different processes, in the atmosphere. The different atmospheric processes can be understood as a function of the pressure ($P$) in the atmosphere. Deep in the atmosphere ($P$ $\gtrsim$ 1 bar) the pressure and temperature, and hence density and opacity, are high enough that thermochemical equilibrium and radiative-convective equilibrium prevail, i.e., chemical reactions occur faster than kinetic processes. The resulting composition then is that which minimises the Gibbs free energy of the system for the given temperature, pressure, and elemental abundances. Higher up in the atmosphere, between $\sim$1mbar and $\sim$1 bar, various processes become more prevalent including atmospheric dynamics, clouds/hazes, and temperature inversions, as a result of complex interplay between the incident radiation field, chemical composition, and other planetary properties. These processes strongly influence, and are influenced by, the atmospheric chemical composition and temperature structure both of which can be out of equilibrium. Further up in the atmosphere ($P$ $\sim$ 10$^{-6}$ - 10$^{-3}$ bar), the low density and high radiation field cause photochemical processes to govern the atmospheric composition through photodissociation of prominent molecules into their constituent atomic species and formation of new ones. Eventually, at very low pressures, atmospheric escape of atomic species lead to mass loss from the atmosphere. Therefore, each region of the atmosphere provides a window into specific physicochemical processes. At the same time, different chemical species and different regions of the atmosphere are accessible to different parts of the electromagnetic spectrum. Prominent molecular species (e.g. H$_2$O, CO, CO$_2$, CH$_4$, etc) absorb primarily in the infrared due to rovibrational transitions, with the exception of some heavy metal species (TiO, VO, TiH, etc.) which also have strong visible absorption. On the other hand, atomic species absorb primarily in the optical and UV, depending on the excitation and ionisation states. Therefore, while UV observations probe the uppermost regions of the atmosphere where the composition is entirely atomic, the infrared observations probe lower regions of the atmosphere where the composition is primarily molecular, with optical spectra probing intermediate regions. Thus the spectral range of observations governs the region in the atmosphere probed and the corresponding physicochemical properties and processes.

The study of exoplanetary atmospheres has progressed at a tremendous pace in recent years. Just a decade ago, fewer than 25 transiting exoplanets were known and the first directly imaged planets were just being discovered. Observations of exoplanetary atmospheres were still in their infancy. Prior to that, only a handful of atomic species were detected robustly, mostly in the two famous transiting hot Jupiters HD~209458b and HD~189733b, using transmission spectra in the optical and UV obtained with the Hubble Space Telescope (HST) \citep{charbonneau2002,vidal-madjar2003}. The first detections of exoplanetary atmospheres using multi-band space-based infrared photometry and spectrophotometry with Spitzer and HST were being made \cite{charbonneau2008, grillmair2008, swain2008}. The instruments used for these early measurements (such as the HST NICMOS spectrograph and Spitzer photometric instruments) were never designed for exoplanetary observations which require extreme sensitivities (e.g., photometric precisions better than $\sim$10$^{-4}$). As such, the early measurements were intensely debated for their accuracy and several were subsequently revised as the systematics were better understood and analysis methods improved. This in stark contrast to the present day as high confidence and reproducible detections of exoplanetary spectra with HST are routine, as discussed below. At the same time the first attempts for transit spectroscopy from the ground were also being made\citep{redfield2008,sing2009a}. Even so, inferences of molecular absorption and pressure-temperature profiles (e.g. thermal inversions) relied heavily on forward models with solar-like elemental compositions and equilibrium conditions \citep{knutson2008,burrows2008}. Statistical constraints on atmospheric properties of exoplanets using atmospheric retrieval methods were still a dream for the future. On another front, the first thermal phase curves, and brightness maps, were being observed for hot Jupiters using infrared photometry with the Spitzer space telescope \citep{knutson2007}. At the same time, the first three-dimensional atmospheric circulation models were being developed for hot Jupiters \citep{showman2008} to explain the observed thermal phase curves. In summary, about a decade ago major observational facilities were being pushed to their limits to detect spectra of transiting exoplanets. At the same time, concomitant developments in theory were being pursued to explain the extant observations, however sparse. Such were the humble beginnings of atmospheric characterisation of exoplanets a decade ago. 

\begin{figure}[t]
\includegraphics[width=1.0\linewidth]{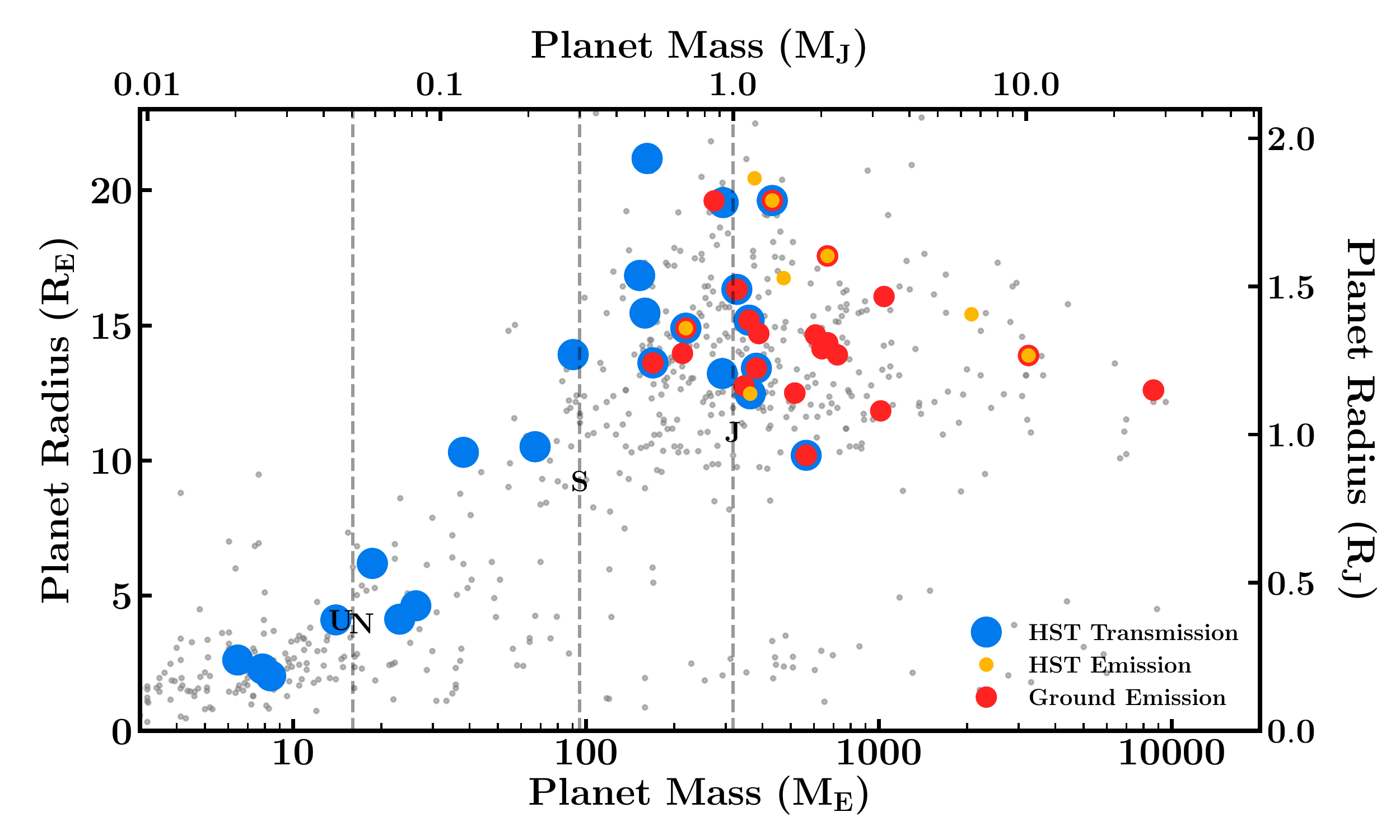}
\caption{Masses versus radii of transiting exoplanets whose atmospheres have been observed. Planets with HST transmission/emission observations are shown by blue/yellow circles, respectively. Emission observations from ground-based facilities are shown in red. All known planets in this mass and radius range, with or without atmospheric observations, are shown for reference as grey points. Solar System planets are denoted in black by their initials and represented by the vertical dashed lines. Data obtained from the NASA Exoplanet Archive at https://exoplanetarchive.ipac.caltech.edu/.}
\label{fig:mass_radius}
\end{figure}

Fast forward to the present - the field has transformed beyond recognition. Firstly, robust detections of exoplanetary atmospheric spectra are routinely made today using a variety of techniques: transit spectroscopy \citep{charbonneau2002}, direct imaging \citep{marois2008}, as well as Doppler spectroscopy \cite{snellen2010}, both from space as well as the ground. Here, robustness of a result implies high statistical significance of a detection and its reproducibility by multiple groups using the same dataset. The space-based instruments used for the purpose in recent years are mainly HST spectrographs such as WFC3 in the near-infrared \citep{mccullough2012,deming2013} and STIS in the optical and UV \citep{sing2016,ehrenreich2015}, and until recently the Spitzer IRAC 1 and 2 bands in the infrared \citep{demory2016}, all of which have been very well characterized compared to the early observations a decade ago. Atmospheres of nearly a hundred exoplanets have been detected using at least one technique in at least one spectral band, as shown in Fig.~\ref{fig:mass_radius}. However, meaningful constraints on atmospheric properties require more than a few photometric observations. The required high-precision spectra over a broad spectral range have been observed for tens of giant exoplanets resulting in robust inferences of their chemical and/or thermal properties. Beyond detecting atmospheric spectra \citep{deming2013}, the advent of atmospheric retrieval techniques \citep{madhusudhan2018} have enabled inverting the observed spectra to obtain detailed statistical estimates of atmospheric properties of exoplanets as a standard procedure. Such observations and inversion techniques are providing initial constraints on key atmospheric properties such as prominent molecular and atomic species (e.g. H$_2$O, CO, CH$_4$, CO$_2$, HCN, TiO, VO, Na, K), elemental ratios (e.g. O/H, C/O ratios), temperature profiles (including thermal inversions), clouds/hazes, circulation patterns, and exospheres. 

The field is now moving beyond atmospheric characterisation of individual exoplanets towards comparative characterisation of ensembles of planets. Such constraints on atmospheric properties of exoplanets have advanced theories of a vast range of corresponding physicochemical processes in exoplanetary atmospheres, spanning chemical and radiative processes, atmospheric dynamics, atmospheric escape, and clouds/hazes. The derived chemical abundances are also being used to investigate constraints on planetary formation and evolutionary processes, which is one of the major current frontiers of the field \citep{madhu2014_migration,lammer2018}. At the same time, the first observations of chemical signatures in atmospheres of low-mass exoplanets, e.g. Neptunes and super-Earths, are becoming feasible, giving rise to the fledgling area of exogeology \citep{tsiaras2016b,esteves2017,wakeford2017}. Observations have also been attempted for transiting planets in the habitable zones of low-mass stars. While no such detection has yet been feasible the prospect of detecting an atmospheric signature of a rocky planet, potentially in the habitable zone, seems like a realisable dream. 

Developments on the observational front have been led by both demonstrations of new detection methods as well as new instrumentation. The most remarkable of these successes have been in three directions corresponding to three detection methods. First, extensive high-precision transit spectroscopy with HST instruments from the UV to NIR have led to atmospheric characterisations in a large sample of transiting exoplanets. In particular, the HST WFC3 spectrograph in the NIR has made H$_2$O detections in exoplanetary atmospheres a routine matter today, with observed planets ranging from dozens of hot Jupiters to exo-Neptunes and even a super-Earth. This is remarkable considering that the H$_2$O abundances are not known for the giant planets in our own solar system owing to their low temperatures at which H$_2$O condenses. Besides HST, remarkable developments have also been made in transit spectroscopy using ground-based facilities. Such observations have led to detections of key species such as Na, K, and TiO, as well as of thermal emission from transiting exoplanets. Second, starting with the first detections a decade ago, direct imaging and spectroscopy of exoplanets is a highly successful technique today with nearly ten objects discovered and various dedicated ground-based surveys now underway. Finally, within this decade high-resolution Doppler spectroscopy in the near-infrared has made it possible to detect molecules in exoplanetary atmospheres, both transiting and non-transiting, by cross-correlation with template spectra. This technique has been very successful in detecting prominent molecules in several giant exoplanets orbiting bright stars. Beyond these broad developments, numerous advances have been made in various aspects of each detection method as will be discussed in this review. Concomitant advancements have also been made in atmospheric modelling, retrieval and theoretical studies. 

The present review is an attempt to discuss the state-of-the-art of this exciting frontier. We will briefly review the recent advances in observational and theoretical methods in sections~\ref{sec:obs_methods} and \ref{sec:theory_advancements}. We will then review the state-of-the-art, challenges, and future landscape of three frontier topics in the area. We will discuss advances in atmospheric characterisation of exoplanets in section~\ref{sec:atmos_characterisation}, those on implications for planetary formation in section~\ref{sec:planet_formation}, and on habitability and biosignatures in section~\ref{sec:habitability}. We will conclude with a discussion on the landscape for the immediate future.

\section{Observational Methods}
\label{sec:obs_methods}

Exoplanetary atmospheres have been observed using a wide range of methods that allow complementary constraints on their physicochemical properties. The methods fall into three broad categories: (a) Transit spectroscopy, (b) High-resolution Doppler spectroscopy, and (c) Direct imaging spectroscopy. While transit spectroscopy and Doppler spectroscopy are most conducive for atmospheric characterisation of close-in planets, direct imaging is more suited for planets at larger orbital separations.  These various methods have been discussed in detail in the literature \citep{seager2010,fischer2014,madhusudhan2014a,kreidberg2018b,birkby2018}. The bulk properties of exoplanets whose atmospheres have been observed are shown in Fig.~\ref{fig:mass_radius}. Here we briefly outline each of these methods, their capabilities and limitations, and their future prospects.  

\begin{marginnote}[120pt]
\entry{HST}{Hubble Space Telescope}
\entry{Spitzer}{Spitzer Space Telescope}
\entry{JWST}{James Webb Space Telescope}
\entry{VLT}{Very Large Telescope}
\entry{GTC}{Gran Telescopio Canarias}
\entry{E-ELT}{European Extremely Large Telescope}
\entry{GMT}{Giant Magellan Telescope}
\entry{TMT}{Thirty Meter Telescope}
\entry{WFIRST}{Wide Field Infrared Survey Telescope}
\entry{HabEx}{Habitable Exoplanet Observatory }
\entry{LUVOIR}{Large UV Optical Infrared Surveyor}
\end{marginnote}

\subsection{Transit Spectroscopy} 

\begin{figure}[h]
\centering
	\begin{subfigure}[b]{0.495\textwidth}
  		\includegraphics[width=\textwidth]{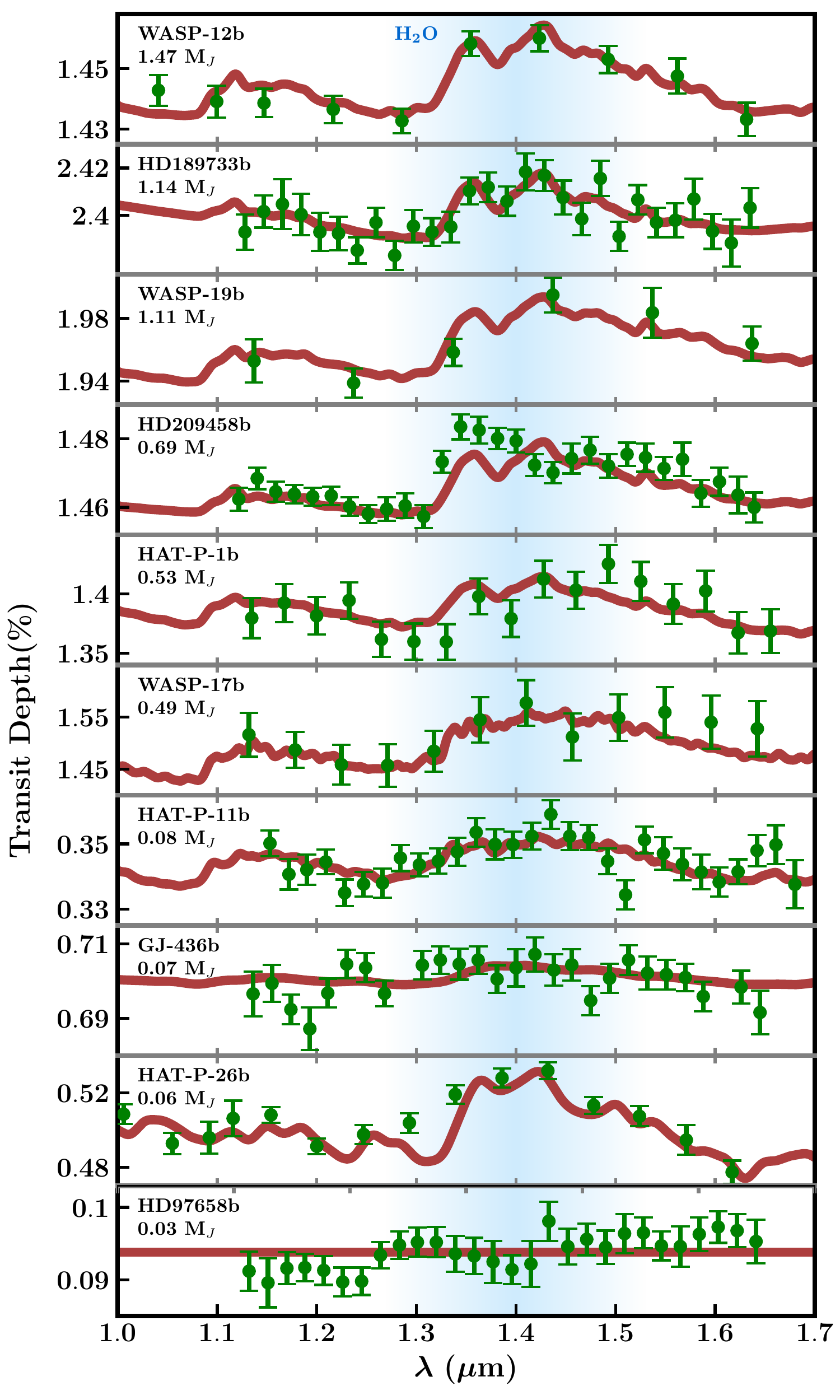}
    \end{subfigure}
	\begin{subfigure}[b]{0.495\textwidth}
  		\includegraphics[width=\textwidth]{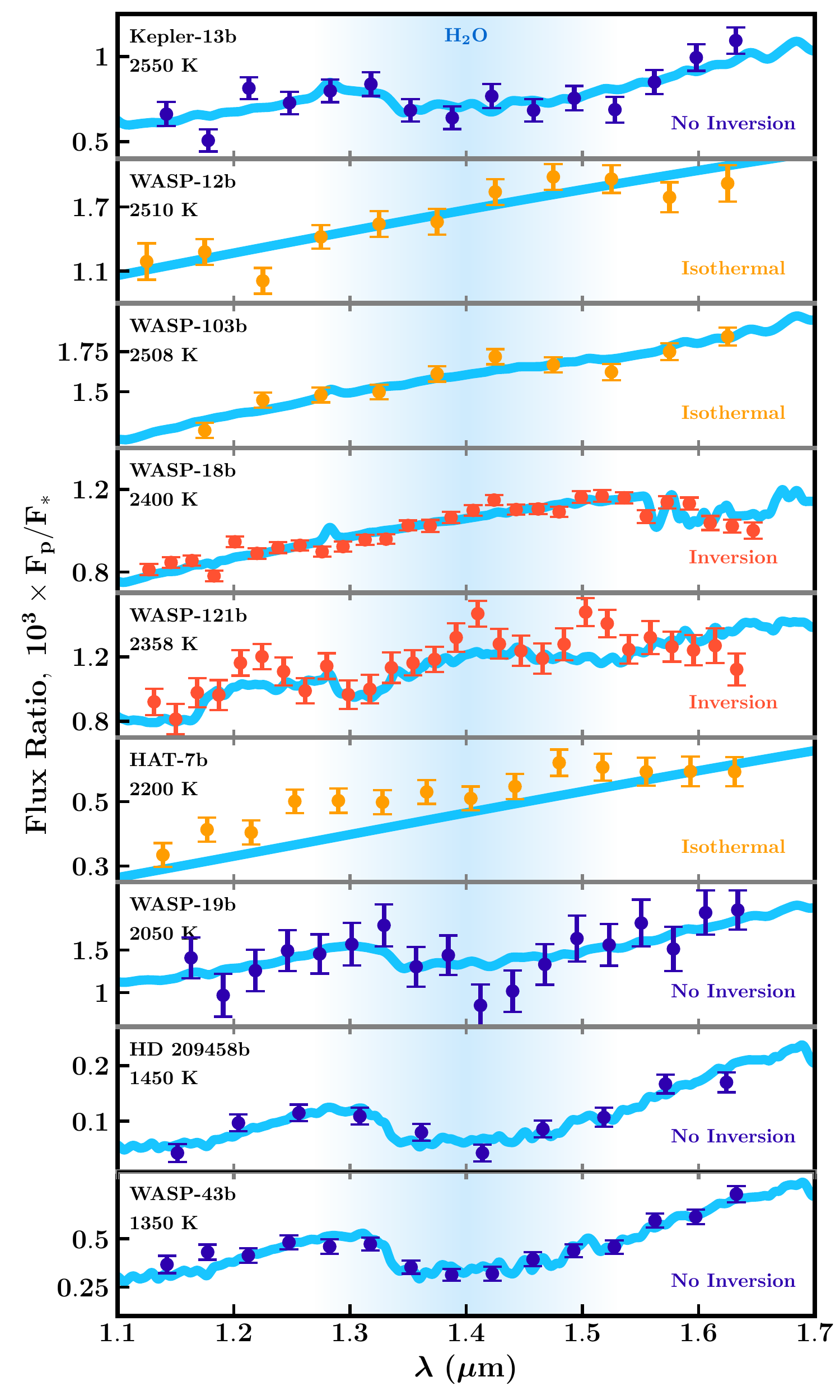}
    \end{subfigure}
\caption{A panorama of primary (left) and secondary (right) eclipse spectra of transiting planets. Observations and error bars are obtained from the sources listed in table \ref{tab:chem_detections}, while nominal models which fit these observations are shown by brown and blue lines. The spectral range affected by water absorption is shaded in blue and water features are present in many of the spectra. Left panel: Transmission spectra of planets arranged by ascending mass. Right panel: Secondary eclipse spectra arranged by ascending temperature and spanning a range of temperature structures from isotherms to profiles with thermal inversions.}
\label{fig:primary_secondary}
\end{figure}

Transit spectroscopy has been the most successful avenue for  their atmospheric characterisation of exoplanets to date, both by number of planets observed and the range of atmospheric constraints obtained \citep{charbonneau2002,deming2013}. This is due to both the larger number of planets detected using the transit method as well as the favourable geometry which makes it relatively easier for atmospheric observations compared to other methods. The transit method allows three configurations to observe a planet's atmosphere: (a) a `transmission spectrum' when the planet transits in front of the host star, i.e. at primary eclipse (b) an emission spectrum as the planet passes behind the host star, i.e. at secondary eclipse, and (c) a phase curve as the planet orbits between the primary and secondary eclipses \citep{knutson2009,stevenson2014}. During the primary eclipse, the star light along the line of sight passes through the atmosphere at the day-night terminator of the planet. The resultant spectrum observed contains absorption features of the planetary atmosphere imprinted on the stellar spectrum. The difference between the in-transit and out-of-transit spectrum, normalised by the out-of-transit spectrum, yields the transmission spectrum. The transmission spectrum is effectively a measure of extinction due to the planetary atmosphere at its day-night terminator region. On the other hand, the secondary eclipse spectrum measures the emergent spectrum from the dayside atmosphere of the planet. Just prior to secondary eclipse the combined spectrum of both the star and the planetary dayside is observed. This combined spectrum, when subtracted by stellar spectrum, which is observed during secondary eclipse, yields the planetary spectrum. Finally, the phase curve provides a spectrum of the planet at different phases. Each of these configurations of transit spectroscopy provides different and complementary constraints on the atmospheric properties of a transiting planet. A panorama of state-of-the-art atmospheric spectra observed during primary and secondary eclipse is shown in Fig. \ref{fig:primary_secondary}. 

A transmission spectrum is essentially a measure of the thickness of the atmosphere probed perpendicular to the line of sight as a function of wavelength. It provides constraints primarily on the chemical composition of the atmosphere at the day-night terminator region, along with the mean molecular weight and temperature through the scale height. Different spectral regions provide constraints on different chemical species. Prominent molecules expected in giant exoplanetary atmospheres such as H$_2$O, CO, CH$_4$, CO$_2$, HCN, TiO/VO, etc., have significant abundances and strong absorption features in the infrared and/or visible wavelengths making them detectable in transmission spectra \citep{madhusudhan2012,moses2013a}. Similarly, atomic features of alkali metals Na and K have strong absorption features in the visible \citep{seager2000,sing2016}. On the other hand, transmission spectra are also excellent probes of scattering in atmospheres. Different sources of scattering such as Rayleigh scattering for small particles versus Mie scattering for larger particles imprint distinct features in the optical transmission spectra and, hence, provide constraints on the presence of clouds/hazes in the atmospheres \citep{wakeford2015}. Furthermore, the presence of high-altitude cloud decks in the atmosphere can also reduce the observable region of the atmosphere and mute the spectral features. Transmission spectra also provide unique probes of exoplanetary exospheres via detections of ionic species which have strong absorption features in the visible and UV \citep{ehrenreich2015}. Recent studies have also investigated the impact of stellar heterogeneity on the spectral features observed in transmission spectra \citep{mccullough2014,oshagh2014}. This is of particular significance to low-mass exoplanets orbiting M Dwarfs which are known to be active \citep{rackham2018}.

An emission spectrum, on the other hand, directly probes the temperature structure of the dayside atmosphere of the planet along with its chemical composition \citep{kreidberg2014b,stevenson2014}. In principle, the full planetary spectrum observed at secondary eclipse contains both reflection and emission. Whereas reflection dominates at optical wavelengths, corresponding to the peak of the stellar spectrum for FGK stars, the planetary emission typically dominates in the infrared owing to the lower temperatures. The planet-star flux contrast increases with wavelength as the star gets fainter and the planet gets brighter with wavelength. Thus, most of the dayside observations of exoplanetary atmospheres have been reported in the infrared. The observed spectrum probes the brightness temperature of the planet at different wavelengths, which effectively translates to measuring the temperatures at different depths in the atmosphere corresponding to the planetary photosphere at different wavelengths. The shapes and amplitudes of the spectral features are governed by both the chemical composition and temperature gradient in the atmosphere. For a given composition, a temperature profile with negative (positive) gradient, i.e.,  temperature decreasing (increasing) with altitude, leads to absorption (emission) features in the emergent spectrum. Therefore, emission spectra can provide powerful constraints on the presence of thermal inversions in exoplanetary atmospheres (discussed further in section~\ref{sec:thermal_inversions}). At the same time, for a given temperature profile the abundances of the chemical species affect the amplitude of the spectral feature. Thus, thermal emission spectra can provide strong constraints on both the composition as well as the temperature profile of the dayside atmosphere. 

A phase curve measures the emergent spectrum of the planet, and hence its atmospheric properties, as a function of orbital phase \citep{stevenson2014}. In addition to the constraints on composition and temperature, thermal phase curves provide direct constraints on the atmospheric dynamics and energy transport in the atmosphere. A thermal phase curve can also be deconstructed to provide the longitudinal temperature distribution in the planetary atmosphere as a function of depth. Thus, atmospheric observations of transiting exoplanets can provide constraints on a wide range of atmospheric processes as discussed above. The specific constraints on the various processes reported by existing observations are discussed in further detail in section~\ref{sec:atmos_characterisation}. 

\subsection{High-resolution Doppler Spectroscopy}
High-resolution Doppler spectroscopy of close-in planets has offered a powerful means to detect chemical species in atmospheres of close-in giant exoplanets, particularly hot Jupiters \citep{snellen2010,birkby2013}. A detailed review of this area can be found in \citep{birkby2018}. This method involves phase-resolved high-resolution (R $\sim10^5$) spectroscopy of the star-planet system to infer the Doppler motion of the planet using the planetary spectral lines. The combined spectra, observed using large ground-based facilities, contain contributions from both the stellar and planetary spectra along with telluric features due to the Earth's atmosphere. For a typical hot Jupiter the RV semi-amplitude of the planet is 1000$\times$ larger than that of the star. Thus, the stellar and telluric features are relatively unchanged during the course of the observations, compared to the planetary spectral lines which undergo significant Doppler shifts. The stellar and telluric features in the data are removed using various detrending methods \citep{brogi2012, birkby2013,cabot2019} which aim to remove the non-varying components leaving behind only the time-varying signal from the planet. In addition, some of the spectral regions with dense telluric contamination are masked out in the data. The resulting residual spectra after detrending are then cross-correlated with template planetary spectra containing the expected prominent molecules. For the matching planet spectrum the orbital motion of the planet can be reconstructed; in particular, the radial velocity semi-amplitude of the planet (K$_p$) and systemic velocity (V$_{\rm sys}$) are constrained. A high significance peak in the K$_p$-V$_{\rm sys}$ plane constitutes a detection of the molecule present in the model template; typically a 5-$\sigma$ signal-to-noise is considered a strong detection. The measured K$_p$, along with the known stellar velocity, also provides an independent constraint on the mass of the planet and the orbital inclination of the system\citep{brogi2012}. This method has been used to detect chemical species in a number of close-in hot Jupiters, e.g. molecular species CO \citep{snellen2010,brogi2012}, H$_2$O \citep{birkby2013}, TiO \citep{nugroho2017}, HCN \citep{hawker2018,cabot2019}, and atomic species Fe, Ti, Ti+ \citep{hoeijmakers2018}. In addition to molecular detections, this technique has also led to constraints on the temperature profiles \citep{nugroho2017} and atmospheric wind speeds\citep{snellen2010,louden2015}. 

 This technique has also been used for atmospheric characterisation of directly-imaged planets at large orbital separations. This has been demonstrated by the CO detection in $\beta$ Pic b  \citep{snellen2014} using a combination of high-contrast imaging and high resolution spectroscopy. Given their large orbital separations such planets are not expected to be tidally locked. Thus, their rotational velocity can be measured via the broadening of the spectral line. A rotational velocity of 25 km/s was measured for $\beta$ Pic b using CO spectral features. The combination of high resolution spectroscopy and high contrast imaging enhances the sensitivities to flux contrasts beyond that achievable by either method. Furthermore, the combination of the cross-correlation technique with adaptive optics assisted integral-field spectrographs at medium resolution can also be used for detecting chemical species across the two-dimensional field of the image. Such `molecular mapping' has been successfully demonstrated for high-significance detections of prominent molecular species such as H$_2$O and CO in directly imaged planets \citep{hoeijmakers2018b, petit2018}. 
 
 The future of high-resolution Doppler spectroscopy offers both exciting avenues and commensurate challenges. With the advent of new high-resolution spectrographs, both on current and future facilities, it is natural to expect this as a promising pathway for molecular detections in exoplanetary atmospheres. Indeed, studies have suggested the feasibility of detecting molecules in the atmospheres of terrestrial-size planets with upcoming large facilities such as the E-ELT \citep{snellen2015,rodler2014}. However, the detection significances of the species are arguably reliant on the metrics used for quantifying the signals and the detrending approaches. Latest studies are beginning to quantify these aspects \citep{cabot2019}. On another front, this technique on its own is less conducive for measuring the abundances of detected species; the signal is sensitive mainly to the line positions rather than the depths. However, the combination of this technique with low-resolution transit spectroscopy where possible may provide a solution in this regard \citep{brogi2017}. A viable future direction is one where new chemical species are first detected using the this technique and the targets, if transiting, are then followed up with low-medium resolution transit spectroscopy for abundance estimates. The recent first inferences of HCN \citep{hawker2018,cabot2019} and atomic species \citep{hoeijmakers2018} in hot Jupiters using this technique may be the first steps in this direction. 

\begin{figure}[t]
\centering
	\begin{subfigure}[b]{0.495\textwidth}
  		\includegraphics[width=\textwidth]{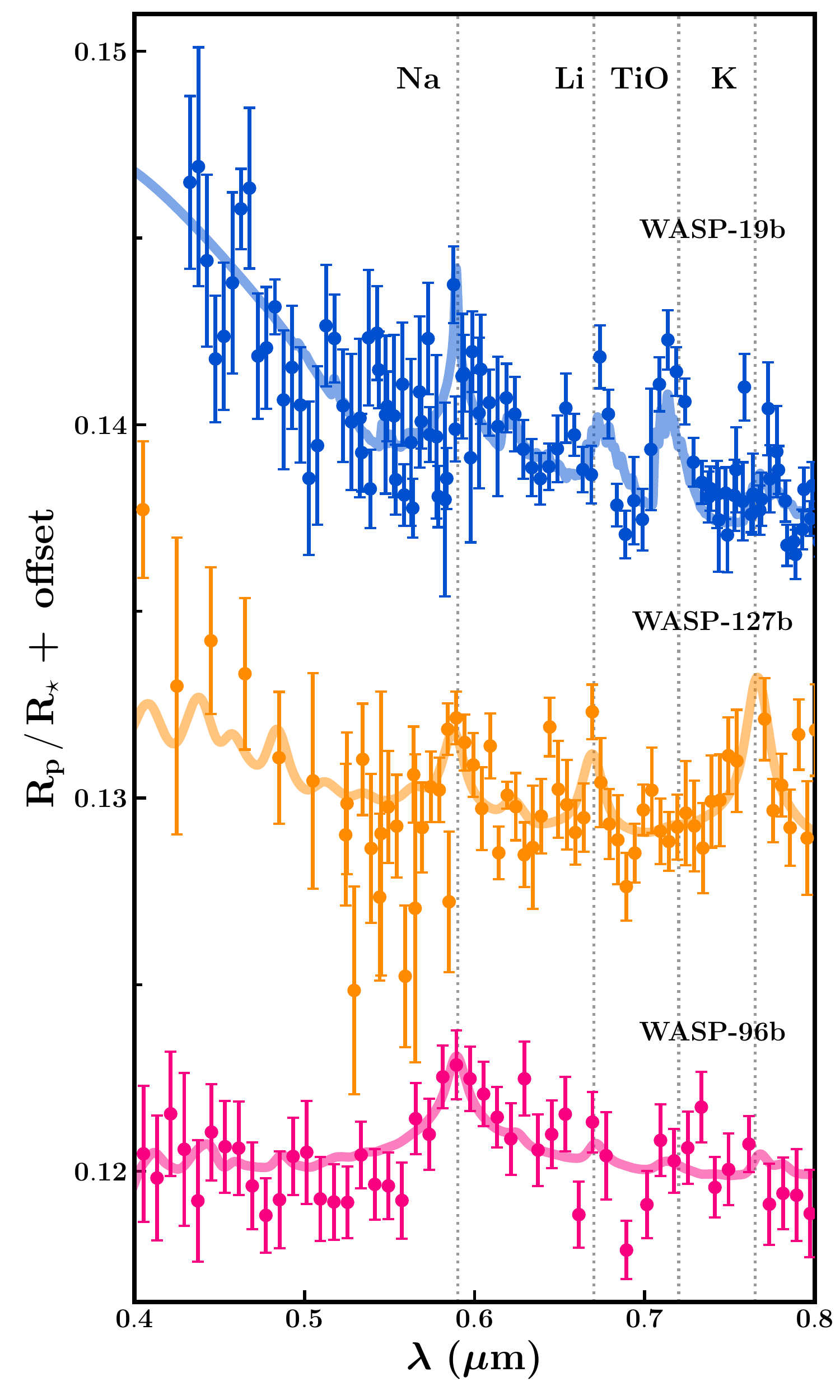}
    \end{subfigure}
	\begin{subfigure}[b]{0.495\textwidth}
  		\includegraphics[width=\textwidth]{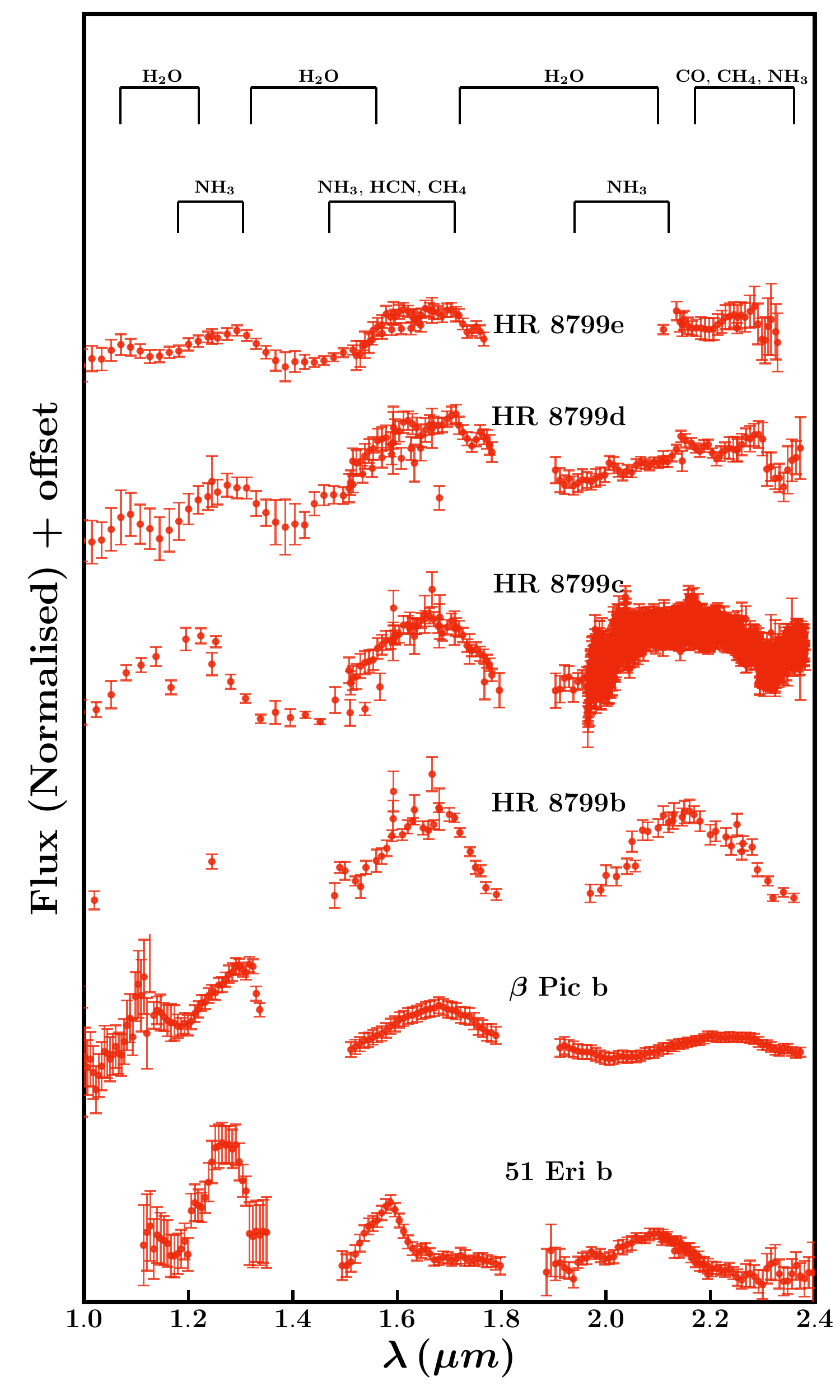}
    \end{subfigure}
\caption{High-quality spectra of transiting (left) and directly-imaged (right) planets obtained with ground-based telescopes (sources are listed in table \ref{tab:chem_detections}). For the spectra of transiting planets, nominal models which fit the data are also shown. The locations of spectral features arising from different chemical species are marked by dashed lines (left) and brackets (right). In both panels, the spectra are arbitrarily offset for clarity.}
\label{fig:di_panorama}
\end{figure}
 
\subsection{Direct Imaging} 

Spectroscopy of planets discovered via direct imaging offers another avenue to characterise exoplanetary atmospheres. In this method, the spectrum of the planet is obtained directly by nulling out the contribution from the star using a coronagraph. Though simple in principle, such observations are challenging given the stringent requirements on the sensitivity and inner-working angle \citep{fischer2014}. For example, a Jupiter analogue orbiting a sun-like star at 10 pc would require a planet-star flux contrast below 10$^{-7}$ in the near-infrared at an inner working angle of 0.5"; the requirements are even more stringent in the optical. However, for young giant planets with high temperatures ($\gtrsim$1000 K) at large orbital separations the planet-star flux contrasts in the near-infrared approaches 10$^{-4}$ making them detectable with current facilities. While the numbers of objects discovered via direct imaging are far fewer than transiting exoplanets, the spectra are typically of higher resolution and higher signal-to-noise ratio owing to the large-aperture ground-based facilities with adaptive optics used for this purpose. As such, the method has been successful in obtaining thermal emission spectra of several young giant exoplanets in the near-infrared. 

The atmospheric properties that can be constrained with directly imaged spectra are similar to those for transiting exoplanets but with some important differences. A directly imaged spectrum is similar to an emission spectrum observed for transiting planets as discussed above and, hence, can provide important constraints on the temperature profile and composition of the atmosphere. However, unlike transiting planets the planetary radius and mass, and hence gravity, are not known a priori. This leads to degeneracies in accurately estimating the chemical compositions from the spectra because the shapes of the spectral features depend strongly on the gravity. Nevertheless, the higher  resolution and signal-to-noise of the observed spectra make it possible to obtain robust detections of chemical species in the  atmospheres notwithstanding the challenges in obtaining specific quantitative constraints on the atmospheric properties. For example, high-confidence detections of H$_2$O, CO, and CH$_4$ have been reported for several directly imaged exoplanets in recent years (see table \ref{tab:chem_detections}). Given the long orbital periods, the spectrum of a planet is typically obtained at a single orbital phase which is unknown, which restricts constraints on atmospheric dynamics. However, precise constraints on the globally-averaged compositions and temperature profiles at the observed phase are possible using atmospheric retrieval techniques. Additionally, given the low irradiation regime the atmospheric temperature profiles of directly imaged planets are expected to be markedly different from those of transiting exoplanets which tend to be highly irradiated. Initial constraints have been reported for a few objects while highlighting the challenges in resolving the various degeneracies. The specific constraints obtained for directly imaged planets are discussed in more detail in section~\ref{sec:atmos_characterisation}.  

The observational landscape of atmospheric characterisation of directly imaged planets is promising, limited only by sample size. Only $\sim$10 directly imaged planets have been discovered since the first detections a decade ago \citep{marois2008,kalas2008}. The small current sample is arguably due to the paucity of giant exoplanets orbiting young stars at large orbital separations with flux contrasts above the detection thresholds of extant surveys. However, with new surveys increasingly aiming at higher sensitivities and inner working angles the sample size is likely to increase \citep{greenbaum2018}. On the other hand, current facilities have already provided spectacular spectra for some of these known planets, such as those in the HR 8799 system \citep[e.g.,][]{barman2011a,konopacky2013,barman2015, marois2008, lee2013, lavie2017,greenbaum2018} and 51 Eri b \citep{macintosh2015}. The high-quality spectra are obtained thanks to high-contrast instruments on large-aperture ground-based telescopes operating in the near-infrared e.g., Keck \citep{konopacky2013}, GPI on Gemini \citep{greenbaum2018}, SCExAO on Subaru \citep{jovanovic2015}, SPHERE on VLT \citep{bonnefoy2016}. In the near future, the JWST will provide a high-stability platform for high-contrast imaging and spectroscopy in the near-mid infrared from space \citep{beichman2017}. In the late 2020s, NASA's WFIRST mission is expected provide another space-based platform for direct imaging of giant exoplanets, particularly in the optical. 

These facilities will be followed by the 25-40 m class telescopes in the next decade (mid-late 2020s), such as the E-ELT (39.3m), TMT (30m), and GMT (25.4 m), which will have unprecedented sensitivity for direct imaging. The E-ELT could in principle allow the detection of some atmospheric signatures of habitable-zone super-Earths and Earth-like planets orbiting the nearest stars \citep{snellen2015}. The high sensitivities required for such observations may be achieved by a combination of high dispersion spectroscopy and high contrast imaging, as has been demonstrated for the giant planet $\beta$ Pic b with the VLT\citep{snellen2014}. While an exciting possibility, such observations will still be limited to spectral regions with relatively weaker telluric contamination and/or thermal background and, hence, limited molecular features in the NIR. On the other hand, space based facilities can provide the capabilities to surmount these limitations by offering broad spectral coverage (UV to IR) and very high sensitivities. Studies are underway for future large space-borne facilities in the 2030s focused on direct imaging and spectroscopy of habitable-zone exoplanets in search of potential biosignatures. Two such examples are HabEx \citep{gaudi2018} and LUVOIR \citep{luvoir2018}. These mission concepts aim at very high contrast $\lesssim$10${^-10}$ direct imaging and spectroscopy over the UV-NIR spectral range using different flux suppression techniques and apertures, e.g. 4-m aperture for HabEx with coronograph and starshade, and 8-12 m aperture for LUVOIR with a coronograph. 

\begin{summary}[SUMMARY POINTS]
\begin{enumerate}
\item Advances in atmospheric spectroscopy have been made in three directions: (1) Transit spectroscopy, (2) Direct imaging, (3) High-resolution Doppler spectroscopy
\item Transit spectroscopy has been the most successful method: Nearly 100 exoplanets with atmospheres detected and over 20 giant exoplanets with high-precision multi-band spectra. 
\item Transit spectroscopy allows observations of transmission spectra of day-night terminator, thermal emission spectra of dayside, and phase curves over the orbit. 
\item Direct imaging provides high S/N thermal emission spectra at a single phase. Nearly 10 directly imaged planets known with high quality spectra available for several of them. 
\item High-resolution Doppler spectroscopy allows detection of chemical signatures in planetary spectra Doppler shifted due to radial velocity of the planet. Chemical detections made in seven hot Jupiters. 
\end{enumerate}
\end{summary}

\section{Theoretical Advancements} 
\label{sec:theory_advancements}

Alongside observations, important advancements are being made in theoretical modelling and inverse methods to investigate exoplanetary atmospheres. These developments can be classified into three main categories: (1) Forward spectral modelling, (2) Retrieval methods (or inverse methods), and (3) Atmospheric theory. The primary observable in the characterisation of an exoplanetary atmosphere is an atmospheric spectrum. Firstly, even before an observation is made, a theoretical model spectrum is required to assess the feasibility of the observation and to predict the potential science return from the observation. This is the goal of forward spectral models which are used to compute spectra of exoplanetary atmospheres under specific assumptions about the atmospheric properties such as chemical abundances, chemical equilibrium, and/or radiative-convective equilibrium. Once the true spectrum of the planet is observed it may or may not match the spectrum predicted a priori. Thus, in practice, the observed spectrum is interpreted using atmospheric retrieval methods, or inverse methods, which involve deriving statistical constraints on the atmospheric properties of a planet from the spectral data using robust parameter estimation methods. The models used in retrievals, in this case, do not assume chemical/radiative equilibrium but rather use parametric atmospheric properties to be constrained by the data. Beyond spectral models and inverse methods, a wide range of studies use detailed theoretical models to investigate the various physical and chemical processes possible in exoplanetary atmospheres, e.g., non-equilibrium chemistry, atmospheric circulation, clouds/hazes, atmospheric escape, thermal inversions, and the like. In what follows, we review the key advancements in these areas. 

\subsection{Self-consistent Models}
Self-consistent models are used to compute spectra of exoplanetary atmospheres for given assumptions about macroscopic parameters such as gravity, irradiation, and elemental abundances. Self-consistent models currently used in the field range from plane-parallel 1D models in chemical and radiative-convective equilibrium to full three-dimensional general circulation models (GCMs). Detailed reviews of such models can be found in various recent works  \citep{madhusudhan2014a,gandhi2017,hubeny2017,heng2017,marley2015}. Here we briefly summarise the state-of-the-art. While 1-D models are the most commonly used to predict and interpret individual spectra, GCMs are used extensively in interpreting phase-resolved spectra and photometric phase curves. 

\subsubsection{1-D Equilibrium Models} 
A one-dimensional self-consistent model typically assumes an elemental composition (e.g., solar abundances) and equilibrium conditions such as thermochemical equilibrium and radiative-convective equilibrium, in a plane-parallel geometry. While the assumption of chemical equilibrium allows computation of the chemical abundances (i.e. atomic or molecular abundances) from the elemental abundances, the condition of radiative-convective equilibrium allows computation of the pressure-temperature ($P$-$T$) profile in the atmosphere consistent with the chemistry. The assumption of radiative-convective equilibrium is particularly relevant for modelling thermal emission spectra of exoplanets because the temperature gradients in the atmosphere play a critical role in the formation of emergent spectral features. The models compute radiative transfer through the atmosphere with the resultant chemical composition and $P$-$T$ profile to generate the spectrum. A number of such models are available in the literature today with varying degrees of sophistication and flexibility. Early models in the field until recently were generally adapted from stellar spectral models and were computed on fixed grids of opacities, e.g., assuming solar elemental ratios with varying metallicities \citep{seager1998,barman2005,fortney2008,burrows2008}. Newer models are now being custom-built for exoplanetary atmospheres and allow more flexible computations of spectra over a wider range of conditions to reflect the possible diversity of exoplanetary atmospheres \citep{drummond2016,molliere2015,malik2017,gandhi2017}. These models span a wide range in chemical abundances (e.g. C/O ratios and  metallicities), irradiation (from highly irradiated to non-irradiated), treatment of clouds/hazes, and strong visible absorbers such as TiO/VO that can cause thermal inversions. These models differ in their treatment of various aspects such as the radiative transfer, radiative versus convective energy transport, scattering, and opacities \citep{hubeny2017}. 

Central to all atmospheric models are the sources of opacity considered. The opacity is driven by the combination of chemical abundances and their absorption cross sections. While the chemical abundances depend on the modelling approach as discussed in this section (e.g. assuming chemical equilibrium,  disequilibrium or parametric abundances), their absorption cross sections are fixed in the models. Therefore, the accuracy of model spectra are critically reliant on the accuracy of the absorption line lists from which the cross sections are derived. Such accurate line lists are required for a wide range of temperatures ($\sim$300 - 4000 K) and compositions possible in exoplanetary atmospheres, which extend beyond traditional applications. This need has been widely recognised in the field and substantial progress has been made in recent years towards accurate high-temperature linelists of numerous molecular species of importance for exoplanetary atmospheres \citep{tennyson2018}. For example, recent line lists are now available for various species including H$_2$O \citep{barber2006, rothman2010}, CO \citep{rothman2010, li2015}, CO$_2$ \citep{rothman2010, huang2013,huang2017}, CH$_4$ \citep{yurchenko2013,yurchenko&tennyson2014}, NH$_3$ \citep{yurchenko2011}, HCN \citep{harris2006,barber2014}, collision-induced-absorption \citep{richard2012} and AlO \citep{patrascu2015}. These developments have greatly improved the accuracies of such line lists as well as their applicability to exoplanetary atmospheres.

\subsubsection{General Circulation Models}
GCMs solve the full three-dimensional structure of the atmosphere given the planetary bulk parameters and irradiation field. These models compute the chemical, thermal, dynamical and radiative properties of the atmosphere in extensive detail. Starting with the first coupled GCMs, with both radiative and dynamical treatment, about a decade ago \citep{showman2009} a number of GCMs are prevalent in the field today with varied levels of complexity. The latest GCMs span a wide range of irradiation conditions, from highly irradiated planets to non-irradiated isolated sub-stellar objects, and orbital parameters (e.g., eccentricities, obliquities), and masses  \citep[e.g.][]{showman2013,parmentier2013,kataria2016,lewis2017,rauscher2017,lines2018,drummond2018,boutle2017,dixon-dobbs2018,medonca2016}. These models have been used to explain phase-resolved spectra and thermal phase curves of hot Jupiters \citep{showman2009, kataria2016,lines2018} and to explore various physical processes in detail \citep{showman2009, parmentier2013,parmentier2016, drummond2018}. One of the key successes of GCMs is the predictions of strong equatorial jets in irradiated hot Jupiters that can lead to shifting of the hot spot in the dayside atmospheres away from the sub-stellar point \citep{showman2002}. This effect, predicted nearly a decade ago \citep{showman2002,showman2009}, has been substantiated by other GCMs in the literature \citep[e.g.][]{rauscher2010,showman2011, drummond2018} as well as by numerous observations of hot Jupiters as discussed in section \ref{sec:dynamics}. These models have also recently considered the effect of latent heat on the atmospheric properties through condensate formation and chemical recombination \citep{tan2017, bell2018}. GCM models also show a trend of day-night temperature contrasts in hot Jupiters increasing with equilibrium temperatures \citep{komacek2017}, consistent with observations and empirical studies \citep{komacek2017,schwartz2017}. Atmospheres of highly irradiated and tidally locked hot Jupiters have also been predicted to contain larger planetary-scale bands compared to those of weakly irradiated and faster rotating planets like the solar-system giant planets. 

Recent studies are moving beyond hot Jupiter atmospheres with solar abundances.
Early GCMs typically assumed solar elemental abundances for hot Jupiters, from which the molecular compositions were calculated assuming thermochemical equilibrium. Recent GCMs are now being used to explore the effect of chemical composition on the dynamical processes in the atmospheres. New GCMs are capable of simultaneously modelling the dynamics coupled with chemistry, radiative transfer, and clouds \citep{parmentier2013,parmentier2016,lewis2017,drummond2018,lines2018}. GCMs are now also exploring lower-mass planets which are within the reach of current and forthcoming observational facilities. In particular, several studies have explored the effect of metallicities in sub-Jovian mass planets such as Neptunes and super-Earths which are not necessarily hydrogen-rich \citep{lewis2010, zhang2017,drummond2018}. Such studies have sought to model and explain observations of thermal phase curves of super-Earths, e.g 55 Cancri e, that are already feasible with current facilities \cite{demory2016}. 

\subsection{Atmospheric Retrieval}
Atmospheric retrieval refers to deriving the atmospheric properties of a planet from its observed spectrum. The canonical retrieval method comprises a parametric forward model of an exoplanetary atmosphere coupled with a parameter estimation algorithm to estimate the model parameters given a spectral dataset. The free parameters in the model include the dominant chemical species with strong features in the observed spectral bandpass, the temperature profile, and the macroscopic clouds/haze parameters, e.g. the location in the atmosphere, spatial extent and opacity, besides any other free parameters relevant to the spectrum at hand. Starting with the first retrieval codes a decade ago, a number of codes are currently available in the field with applicability over the wide range of observations possible. A recent review of existing retrieval codes and capabilities in the field can be found in \citep{madhusudhan2018}. The latest developments in retrievals include incorporation of non-equilibrium phenomena at varying levels of complexity, such as different prescriptions for clouds/hazes \citep[e.g.][]{barstow2017,line2016,lavie2017,macdonald2017}, deviations from radiative-convective equilibrium \citep{gandhi2017}, and benchmarking with three-dimensional circulation models \cite{blecic2017}. On the other hand, state-of-the-art codes use a wide range of parameter estimation methods spanning Markov chain Monte Carlo methods \cite{blecic2017}, optimal estimation gradient-descent algorithms \cite{barstow2017}, nested sampling algorithms \citep[e.g.,][]{benneke2013}, and machine learning algorithms \citep{waldmann2016}. In recent years, state-of-the-art retrieval codes have allowed detailed constraints on the chemical compositions, temperature profiles, and properties of clouds/hazes in a number of exoplanets, both transiting and directly-imaged \citep{madhusudhan2018}. 

\subsection{Disequilibrium Models}

Planetary atmospheres are seldom in equilibrium in entirety. Various processes can drive an atmosphere out of chemical, radiative, and thermal equilibria. These processes, illustrated in Fig.~\ref{fig:atm_schematic}, include vertical mixing and photochemical processes, circulation, clouds/hazes, atmospheric escape, etc., one or more of which dominate in any observed region of the atmosphere. The ultimate model atmosphere would include all these processes simultaneously over the entire atmosphere, troposphere to exosphere. However, such a comprehensive unified model for exoplanetary atmospheres is impractical at present. Additionally, exoplanetary spectra are typically observed in a limited spectral band at a time constraining some atmospheric properties/processes but with almost no constraints on other aspects. Therefore, a realisable and useful approach currently is to explore each individual processes in detail while allowing for appropriate boundary conditions or simplified prescriptions to represent the interplay with other processes. This has been the approach in the field. Theoretical studies in recent years have explored each of the aforementioned processes in exoplanetary atmospheres in varied detail as summarised below. 
Here we only discuss briefly developments in atmospheric chemistry and clouds/hazes. A comprehensive review of theoretical developments in atmospheric escape processes can be found in \citep{owen2018}. 

\subsubsection{Atmospheric Chemistry}

State-of-the-art models of atmospheric chemistry in exoplanets span a wide range in complexity. While the most extensive chemical networks are considered in 1-D models of non-equilibrium chemistry \citep[e.g.,][]{moses2013a, venot2014, tsai2017}, recent studies are beginning to combine chemical codes in 3-D General Circulation Models \citep{zhang2017,drummond2018}. A recent review of such models can be found in  \citep{madhusudhan2016}. The chemistry in a planetary atmosphere is a strong function of the macroscopic parameters such as the elemental abundances (e.g., of H, O, C), stellar irradiation, gravity, and mean molecular mass. In particular, there is a clear dichotomy between primary atmospheres which are expected to be dominated by H$_2$/He as in the giant planets and secondary atmospheres that are expected to be dominated by heavier molecules such as H$_2$O, CO$_2$, or N$_2$, as in the terrestrial planets. Given the bulk properties, the key processes governing the chemistry in a planetary atmosphere include chemical equilibrium, mixing processes, photochemistry, and chemical diffusion. Each of these processes dominate in a particular region of the atmosphere (see e.g. Fig.~\ref{fig:atm_schematic}), depending primarily on the incident irradiation and, hence, the temperature profile in the atmosphere. Chemical equilibrium dominates in the deep atmosphere, typically for $P\gtrsim1$ bar where the high density and temperature lead to fast thermochemical reactions. At the other extreme, photochemical reactions dominate in the upper atmosphere typically for $P\lesssim10^{-3}$ bar where the incident UV/optical flux is high and the densities are too low for thermochemical reactions to dominate. In the intermediate regions between these two extremes dynamical processes, such as vertical mixing, dominate the chemical composition of the atmosphere. Recent studies have investigated the critical dependence of all aspects of chemistry in exoplanetary atmospheres on the macroscopic parameters, e.g. the level of stellar irradiation, metallicity, and C/O ratio \citep{moses2013a, moses2014, venot2015, zhang2017, drummond2016}.

\subsubsection{Clouds/Hazes}

A critical factor in the understanding of exoplanetary atmospheres is the prevalence of clouds/hazes which can affect both the atmospheric processes and the observed spectra. Numerous theoretical studies have posited the ubiquity of clouds in atmospheres at all temperatures. Detailed reviews on models of clouds/hazes in exoplanetary and sub-stellar atmospheres can be found in \cite{marley2015,madhusudhan2016}. While the clouds in solar system planets are composed of volatile-rich condensates (e.g. H$_2$O, NH$_3$, hydrocarbons, etc.), those in hot exoplanets can span a wide range of refractory-rich compositions. Models of exoplanetary atmospheres over the years have explored the effects of a wide range of clouds/hazes on exoplanetary spectra, both for irradiated planets observed via transit as well as directly-imaged planets. Recent developments on the understanding of clouds/hazes in exoplanetary atmospheres have proceeded in two directions. On one hand, there have been major efforts on self-consistent cloud modelling at varying levels of detail - from one-dimensional self-consistent models \citep[e.g.,][]{marley2013,molliere2015,drummond2016} to three dimensional cloud models \citep{parmentier2013,drummond2018}, as well as detailed studies of clouds microphysics \citep{helling2016}. On the other hand, various studies have explored the range of cloud compositions observable with existing and upcoming facilities \citep{mbarek2016,wakeford2015, pinhas2017}, spanning from refractory clouds in high-temperature atmospheres to volatile clouds in low-temperature atmospheres.   

\begin{summary}[SUMMARY POINTS]
\begin{enumerate}
\item Advancements in atmospheric modelling of exoplanets have been made in various directions, spanning 1-D to 3-D self-consistent models, atmospheric retrievals, and models of various disequilibrium processes. 
\item Self-consistent spectral models are used to compute the spectra of exoplanetary atmospheres given the macroscopic parameters (e.g., gravity, irradiation, elemental abundances) and assumptions of chemical and/or radiative-convective equilibrium. 
\item Forward models range from one-dimensional equilibrium models to three-dimensional general circulation models (GCM).
\item Atmospheric retrievals, or inverse modelling, involves using parametric models and parameter estimation methods to derive atmospheric properties from observed spectra. Derived properties include chemical compositions, temperature profiles, cloud/hazes, and deviations from chemical or radiative equilibrium. 
\item Disequilibrium models include detailed modelling of various processes that drive atmospheres out of chemical and/or radiative equilibria, e.g., kinetic processes, photochemistry, clouds/hazes, and atmospheric escape. 
\end{enumerate}
\end{summary}

\section{Atmospheric Characterisation of Exoplanets}\label{sec:atmos_characterisation}
The combination of state-of-the-art spectroscopic observations and theoretical modelling and retrieval techniques have led to detailed constraints on a wide range of atmospheric properties in numerous exoplanets. Atmospheres of nearly one hundred exoplanets have now been observed in at least two photometric bandpasses, low resolution spectra have been obtained for nearly 40 planets, and medium-high resolution spectra obtained for nearly ten exoplanets. This is a revolutionary development from a decade ago when barely 25 exoplanets were known to transit and the first directly imaged planets were being discovered. These new observations have led to unprecedented constraints on chemical compositions, temperature profiles, clouds/hazes, atmospheric dynamics, and atmospheric escape, for numerous exoplanets. Here we highlight some of the most recent advances in the characterisation of exoplanetary atmospheres. 

\subsection{Chemical Compositions} 
\label{chem_compositions}
The wide spectral range of exoplanetary spectra observed has enabled robust detections of several key chemical species in their atmospheres, as shown in Table~\ref{tab:chem_detections}. As alluded to in section \ref{sec:intro}, atomic and ionic species have strong absorption in the UV and visible due to their electronic transitions. On the other hand, prominent molecular species of volatile elements such as H$_2$O, CO, CH$_4$, etc. show strong absorption in the infrared due to their rovibrational transitions, though heavier molecules such as TiO and VO also show strong absorption in the visible. The combination of spectroscopic measurements at different wavelengths and different observing techniques has led to a range of chemical compositions observed in diverse planets. The different chemical species detected probe different regions in the planetary atmosphere, as shown in Fig.~\ref{fig:atm_schematic}. The strong UV and visible absorbers such as the atomic species probe the upper regions of the atmospheres where photochemistry is most active, with the ionic species probing the exospheres. On the other hand, the molecular species probe the infrared photosphere between $\sim$1 mbar - 1 bar. Table~\ref{tab:chem_detections} shows the chemical species detected using each observing method. While transmission spectra have been used to probe atmospheres over the entire spectral range from UV - infrared, the emission spectra of both transiting planets and directly imaged planets have been observed predominantly in the infrared where the thermal emission peaks. 
 
\begin{center}
\begin{longtable}{c c}
\caption{Chemical detections in exoplanetary atmospheres with different observing techniques.\label{tab:chem_detections}}\\
\multicolumn{2}{c}{}\\
\multicolumn{2}{c}{{\large Transmission Spectra (Primary Eclipse)}}\\
\hline
\hline
\rule{0pt}{3ex}  Chemical Species & Planet (References) \\[1ex]
\hline
\\ [-3ex]
\rule{0pt}{3ex}
\endfirsthead
\multicolumn{2}{c}%
{\tablename\ \thetable\ -- \textit{Continued from previous page}} \\
\hline
\rule{0pt}{3ex} 
Chemical Species & Planet (References) \\[1ex]
\hline
\\ [-3ex]
\rule{0pt}{3ex}
\endhead
\hline \multicolumn{2}{r}{\textit{Continued on next page}} \\
\endfoot
\hline
\endlastfoot

H$_2$O & 
 \parbox{8cm}{\centering HD 189733b \cite{mccullough2014}, HD 209458b \cite{deming2013}, \\ \centering WASP-12b \cite{kreidberg2015}, \centering WASP-17b \cite{mandell2013}, \centering WASP-19b \cite{huitson2013}, \centering WASP-39b \cite{wakeford2018}, \centering WASP-43b \cite{kreidberg2014b}, \centering WASP-52b \cite{tsiaras2018}, \centering WASP-63b \cite{kilpatrick2018}, \centering WASP-69b \cite{tsiaras2018}, \centering WASP-76b \cite{tsiaras2018},  \centering WASP-121b \cite{evans2016}, \centering HAT-P-1b \cite{wakeford2013}, \centering HAT-P-11b \cite{fraine2014}, \centering HAT-P-18b \cite{tsiaras2018}, \centering HAT-P-26b \cite{wakeford2017}, \centering HAT-P-32b \cite{damiano2017}, \centering HAT-P-41b \cite{tsiaras2018}, \centering XO-1b \cite{deming2013}} 
\\
\\

 Na 
 &
 \parbox{8cm}{\centering HD 189733b \cite{redfield2008},  \centering HD 209458b \cite{charbonneau2002},  \centering WASP-17b \centering \cite{sing2016},  \centering WASP-39b \cite{nikolov2016},  \centering WASP-52b \cite{chen2017},  \centering WASP-69b \cite{casasayas2017},  \centering WASP-96b \cite{nikolov2018},  \centering WASP-127b \cite{chen2018},  \centering HAT-P-1b \cite{nikolov2014},  \centering XO-2b \cite{sing2012} } 
\\
\\
 
 K 
 &
 \parbox{8cm}{ \centering WASP-6b \cite{nikolov2015},  \centering WASP-31b \cite{sing2015},   \centering  WASP-39b \cite{sing2016},  \centering  WASP-127b \cite{chen2018},  \centering  HAT-P-12b \cite{sing2016},   \centering  XO-2b \cite{sing2011},  \centering HD 80606b \cite{colon2012}} 
 \\
 \\

 TiO 
 &
 \parbox{8cm}{ \centering WASP-19b \cite{sedaghati2017}}
 \\
 \\

 AlO
 &
 \parbox{8cm}{ \centering WASP-33b \cite{vonEssen2019}}
\\
 \\
 
 H 
 &
 \parbox{8cm}{ \centering HD 189733b  \cite{jensen2012,bourrier2013}, \centering HD 209458b  \cite{vidal-madjar2003,jensen2012}, \centering GJ 436b \cite{ehrenreich2015}} 
  \\
\\
 
He 
 &
 \parbox{8cm}{ \centering WASP-107b \cite{spake2018}} 
 \\
 \\

C
 &
 \parbox{8cm}{ \centering HD 209458b \cite{vidal-madjar2004}} 
 \\
 \\

O
 &
 \parbox{8cm}{ \centering HD 209458b \cite{vidal-madjar2004}} 
 \\
 \\

 Li 
 &
 \parbox{8cm}{ \centering WASP-127b \cite{chen2018}} 
 \\
 \\
 
 Ca 
 &
 \parbox{8cm}{ \centering HD 209458b \cite{astudillo-defru2013}} 
 \\
 \\
 
 Sc 
 &
 \parbox{8cm}{ \centering HD 209458b \cite{astudillo-defru2013}} 
\\
\\
 
 Mg 
 &
 \parbox{8cm}{ \centering WASP-107b \cite{vidal-madjar2013}, WASP-12b \cite{fossati2010}} 
\\
\\
 
 Si 
 &
 \parbox{8cm}{ \centering HD 209458b \cite{schlawin2010}} 
 \\
 \\

\hline
\\[2ex]
\multicolumn{2}{c}{{\large Emission Spectra (Secondary Eclipse)}}\\

\hline
\\

 H$_2$O 
 & 
 \parbox{8cm}{ \centering WASP-43b  \cite{kreidberg2014b}, HD 209458b \cite{line2016}, HD 189733b \cite{crouzet2014}, WASP-121b \cite{evans2017}, Kepler-13Ab \cite{beatty2017b}, WASP-33b \cite{haynes2015}}  
  \\
  \\
 
 CO 
 & 
 \parbox{8cm}{ \centering WASP-18b  \cite{sheppard2017}}  
  \\
  \\
 
  VO 
 & 
 \parbox{8cm}{ \centering WASP-121b  \cite{evans2017}}  
  \\
  \\
  
 TiO
 & 
 \parbox{8cm}{ \centering WASP-33b  \cite{haynes2015}}  
  \\
  \\
  
 HCN
 & 
 \parbox{8cm}{ \centering HD 209458b \cite{hawker2018}}  

\\\hline
\\[2ex]
\multicolumn{2}{c}{{\large High-resolution Doppler Spectroscopy}}\\

\hline
\\

 H$_2$O 
 & 
 \parbox{8cm}{ \centering 51 Peg b \cite{birkby2017}, HD 179949 b \cite{brogi2014}, HD 189733b \cite{birkby2013}, HD 209458b \cite{hawker2018}}  
  \\
  \\
 
 CO 
 & 
 \parbox{8cm}{ \centering $\tau$ Bootis b \cite{brogi2012}, HD 209458b \cite{snellen2010}, 51 Peg b \cite{brogi2013}, HD 179949 b \cite{brogi2014}, HD 189733b \cite{rodler2013,brogi2016}}
 \\
 \\
 
 TiO
 & 
 \parbox{8cm}{ \centering WASP-33b  \cite{nugroho2017}}  
  \\
  \\
  
 HCN
 & 
 \parbox{8cm}{ \centering HD 209458b \cite{hawker2018}, HD 189733b \cite{cabot2019}}  
 \\
 \\
 
  Ti, Fe, Ti+
 & 
 \parbox{8cm}{ \centering KELT-9b  \cite{hoeijmakers2018}}  
  \\

\hline
\\[2ex]
\multicolumn{2}{c}{{\large Direct Imaging}}\\

\hline
\\

 H$_2$O 
 & 
 \parbox{8cm}{ \centering HR 8799b  \cite{lee2013}, \centering HR 8799c \cite{konopacky2013}, \centering HR 8799d \cite{lavie2017}, \centering HR 8799e  \cite{lavie2017},  \centering $\kappa$ And b \cite{todorov2016}, \centering 51 Eri b \cite{samland2017}, \centering Gl 570D \cite{line2015}, \centering HD 3651B \cite{line2015}, $\beta$ Pic \cite{chilcote2017}, \centering ULAS 1416 \cite{line2017}}  
  \\
  \\

  CH$_4$ 
 & 
 \parbox{8cm}{ \centering HR 8799b \cite{barman2015}, \centering 51 Eri b \cite{samland2017}, \centering GJ 504 \cite{janson2013}, \centering GJ 758 B \cite{janson2011}, \centering Gl 570D \cite{line2015},  \centering HD 3651B \cite{line2015}, \centering ULAS 1416 \cite{line2017}} 
 \\
 \\

 NH$_3$ 
 & 
 \parbox{8cm}{\centering Gl 570D \cite{line2015},  \centering HD 3651B \cite{line2015}, \centering ULAS 1416 \cite{line2017}} 
 \\
 \\
 
 CO 
 & 
 \parbox{8cm}{ \centering HR 8799b \cite{lee2013}, \centering HR 8799c \cite{konopacky2013}} 
 \\[1ex]

\end{longtable}

\end{center}

\subsubsection{Atomic and Ionic Species}
The first chemical detections in exoplanetary atmospheres were of atomic and ionic species observed using HST transmission spectra in the optical and UV \citep{charbonneau2002,vidal-madjar2003}. Nearly 20 chemical species have now been detected in exoplanetary atmospheres as shown in table~\ref{tab:chem_detections}. In recent years, chemical species have been detected both from space, using HST, as well as large ground-based telescopes. Foremost among these are detections of the alkali species Na and K, one or both of which have now been detected in over a dozen transiting hot Jupiters. The Na doublet has peaks near 576.8 nm whereas K peaks near  778.8 nm. Many of the Na and K detections have been made using the HST STIS spectrograph in the visible \citep[e.g.,][]{sing2016}. Recently, He was also detected in the atmosphere of a giant exoplanet using HST at 1083 nm \citep{spake2018}. A major development in recent years has been the ability to detect these species routinely and robustly using ground-based facilities. While the first ground-based Na/K detections were already made a decade ago \citep[e.g.,][]{redfield2008}, recent observations are demonstrating nearly space-quality spectra from large ground-based telescopes such as VLT and GTC \citep{nikolov2018,chen2018}. Besides Na and K, the first potential inference of Li was also made from the ground recently using the GTC \citep{chen2018}. The above detections were made using medium-resolution spectra from space and the ground. On the other hand, atomic species have also been detected using very high-resolution spectroscopy of transiting planets, e.g. Na and K in the hot Jupiter HD~189733b \citep{wyttenbach2015} and Ti and Fe in the ultra hot Jupiter KELT-9b \citep{hoeijmakers2018}. 

In addition, several atomic and ionic species have also been discovered using transmission spectra in the UV with HST. These are elements in the exosphere which originated from photodissociation of molecules in the lower atmosphere followed by interactions with higher energy UV photons in the upper atmosphere. Several of the detections are of Lyman alpha absorption of exospheric hydrogen (H) at 121.6 nm using the HST STIS instrument \citep{vidal-madjar2003, ehrenreich2015}. On the other hand, some heavier elements such as C, O, and Mg, have also been detected in the NUV \cite{vidal-madjar2004,fossati2010}. Overall, such exospheric elements have been discovered in nearly ten giant exoplanets, mostly for hot Jupiters but also including a hot Neptune \citep{kulow2014, ehrenreich2015}. Efforts to detect exospheric elements in super-Earths have not yet resulted in detections but have provided important upper limits on the composition of the lower atmosphere. For example, the non-detection of H in the super-Earth 55 Cancri e rules out a hydrogen-rich atmosphere in the planet and places an upper-limit on the possible amount of H$_2$O in the atmosphere \citep{ehrenreich2012}. 

\subsubsection{Molecular Species}
The advent of high-sensitivity infrared spectrographs has made it feasible to robustly detect molecules in exoplanetary atmospheres. The most important molecules in planetary atmospheres are those containing the prominent elements after H and He, namely O, C, and N. Theoretical studies have long predicted that key volatile molecules such as H$_2$O, CH$_4$, CO, HCN, CO$_2$ should be present in H$_2$-rich atmospheres at high temperatures, e.g. of hot Jupiters, depending on the metallicity, temperature, and C/O ratios \citep[e.g.,][]{burrows1999,moses2011,madhusudhan2012}. However, the search for these molecules in exoplanetary atmospheres is usually limited by the spectral range and sensitivities of available instruments. A decade ago only a handful of molecular inferences were reported with very low resolution spectro-photometric data which have since been revised in subsequent analyses. Here we focus on the current state-of-the-art molecular detections. 

A major development in this area occurred in the last few years with the advent of the HST Wide Field Camera 3 (WFC3) \citep{mccullough2012} which allowed high-precision near-infrared spectroscopy of transiting exoplanets \citep{deming2013}. The spectral range of the HST WFC3 G141 grism of 1.1-1.7 $\mu$m contains strong spectral bands of key volatile species such as H$_2$O, CH$_4$, NH$_3$, and HCN, making it a prime instrument for molecular spectroscopy of exoplanetary atmospheres. Among these species, H$_2$O has the strongest features in this band and is also predicted to be the most abundant oxygen-bearing species, besides CO, in high temperature atmospheres \citep{madhusudhan2012,moses2013a}. In the last five years, spectra of nearly 40 transiting exoplanets have been observed with HST WFC3 and robust detections of H$_2$O have been reported in over ten transiting exoplanets \citep{deming2013,mccullough2014,kreidberg2014b,sing2016,tsiaras2018}. Most of the detections were made for gas giants, though H$_2$O has also been detected in a few exo-Neptunes \citep{fraine2015,wakeford2017}. Beyond H$_2$O, initial indications have also been suggested for HCN and NH$_3$ using HST WFC3 transit spectroscopy \citep{macdonald2017_Nchem, kilpatrick2018}. In addition, signs of TiO, VO, and AlO have also been reported using HST and/or large ground-based facilities \citep{haynes2015,evans2017,sedaghati2017,vonEssen2019}. 

\begin{figure}[h]

\includegraphics[width=\textwidth]{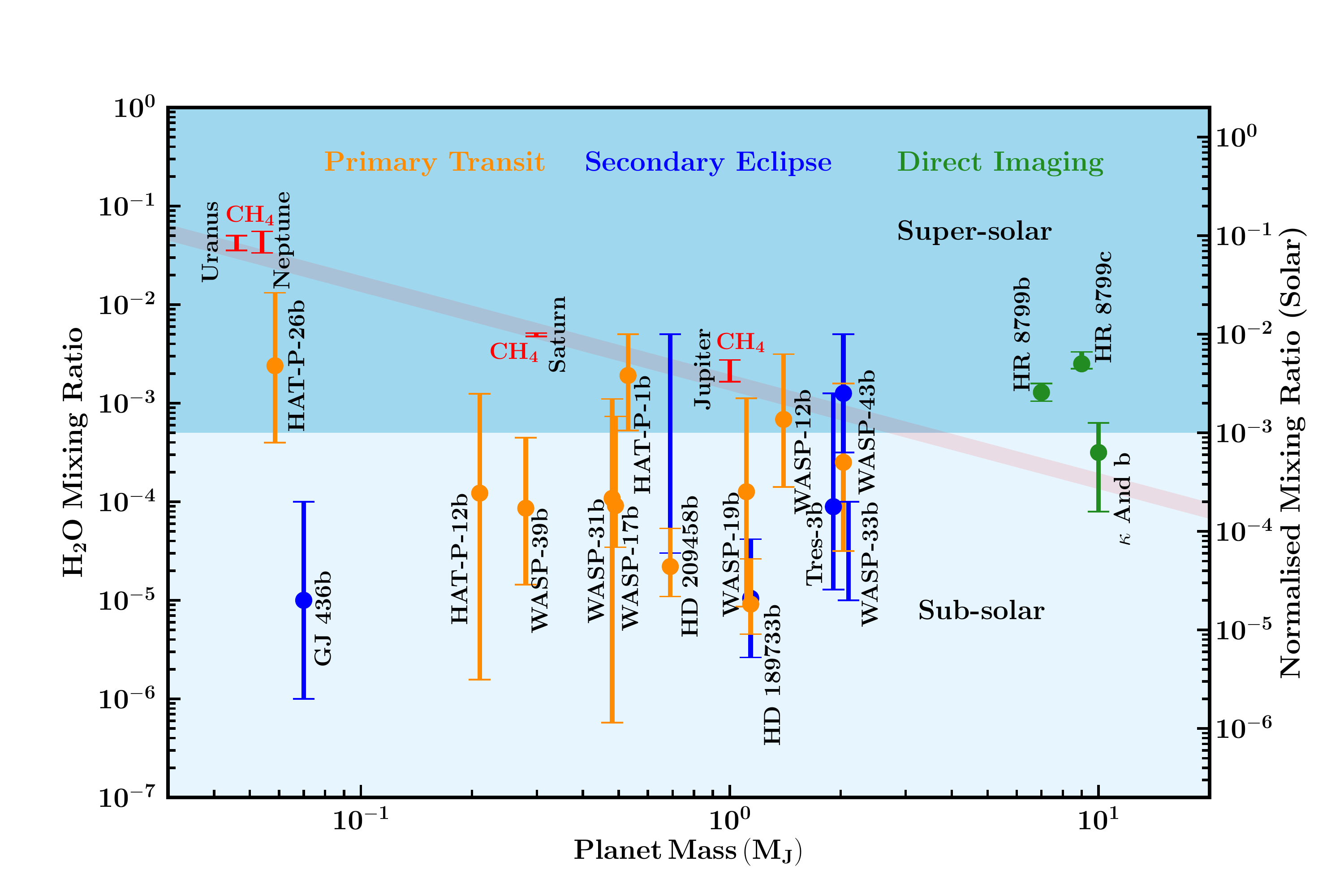}
\caption{Atmospheric H$_2$O abundances for exoplanets in the literature. The abundances shown here are those with uncertainties smaller than 2 dex. The methods used in each case are denoted by colour; primary transit (orange), secondary eclipse (blue) and direct imaging (green). The regions of sub-solar and super-solar abundances are shown in light and dark blue, respectively. Methane abundances of the Solar System planets are shown in red (since their water abundances are not known), and a power-law fit for these measurements is shown by the red line. The methane abundances are obtained from the following sources: Jupiter and Saturn \citep{atreya2016, wong2004, fletcher2009}, Neptune \citep{karkoschka2011}, and Uranus \citep{sromovsky2011}. The exoplanet H$_2$O abundances are from various works: HD 209458b \citep{pinhas2018_h2o, line2016}, HD 189733b \citep{pinhas2018_h2o, waldmann2015a}, WASP-12b \citep{pinhas2018_h2o}, WASP-43b \citep{kreidberg2014b}, WASP-33b \citep{haynes2015}, TrES-3 \citep{line2014}, GJ 436b \citep{moses2013b}, HAT-P-26b \citep{wakeford2017}, WASP-39b \citep{pinhas2018_h2o}, HR 8799 planets \citep{lavie2017}, $\kappa$ And b \citep{todorov2016}.}
\label{fig:h2o_mass}
\end{figure}

Complementary to transmission spectra, molecular detections have also been reported using thermal emission spectra of transiting exoplanets. Such spectra using HST WFC3 have led to detections of H$_2$O in the dayside of several hot Jupiters \citep[e.g.,][]{crouzet2014,kreidberg2014b,line2016} as well as high-temperature molecules such as TiO \citep{haynes2015} and VO \citep{evans2017}. At a lower resolution, inferences of CO have been reported in some hot Jupiters based on infrared photometry using Spitzer in the 3.6 and 4.5 $\mu$m IRAC bands \citep[e.g.,][]{sheppard2017}. 

Beyond transit spectroscopy, more molecular detections at high confidence have now been made through direct imaging and Doppler spectroscopy. High resolution Doppler spectroscopy has led to detections of various molecules including CO \citep{snellen2010,brogi2012}, H$_2$O \citep{birkby2013}, TiO \citep{nugroho2017}, and HCN \citep{hawker2018} in hot Jupiters. At the same time, spectroscopy of directly imaged planets have led to detections of CO \citep{konopacky2013}, H$_2$O \citep{barman2015,konopacky2013,todorov2016,samland2017}, and CH$_4$ \citep{barman2015,samland2017,janson2013} in the atmospheres of several planets. Typically, current data quality for directly imaged objects is significantly better compared to transit planets  \citep{konopacky2013,todorov2016,macintosh2016}. On the other hand, comparative characterisation of directly imaged planets are limited by the much smaller number of objects know, compared to transiting planets, and the lack of prior knowledge about the masses, radii, and gravity.  

\subsubsection{Chemical Abundances}
Beyond detections of chemical species, the high quality spectra have also led to constraints on the atmospheric chemical abundances using detailed retrieval methods. A recent review of retrieval methods and abundance constraints can be found in \citep{madhusudhan2018}. Here, we briefly highlight the state-of-the-art abundance constraints. Despite the detections of a wide range of chemical species in tens of exoplanets, robust abundance constraints are available for only a subset of those planets due to the challenges in retrieving abundances from observed spectra. Firstly, constraints on chemical abundances are presently possible using both transit spectroscopy (transmission and emission) and thermal emission spectra of directly imaged planets. Currently, high resolution Doppler spectroscopy on its own is not as sensitive to abundance determinations \citep{birkby2013,brogi2017}. Secondly, even within transit spectroscopy reliable estimations of chemical abundances require a wide spectral coverage and very high precision observations. In particular, in transmission spectroscopy robust spectra across the visible to near-infrared are required to break the degeneracies between the chemical abundances and the presence of clouds/hazes in the atmosphere. In emission spectra, high-precision observations are required to break the degeneracies between abundances and temperature profiles. 

Currently, the most stringent constraints on chemical abundances have been possible for H$_2$O. Observations of transit spectroscopy in the HST WFC3 band at 1.1-1.7 $\mu$m span a strong H$_2$O feature near 1.4 $\mu$m. Furthermore, the HST STIS band in the optical provides important constraints on clouds/hazes in the atmosphere, thereby resolving degeneracies with composition. Such observations have been used to retrieve H$_2$O abundances for over ten transiting hot Jupiters and Neptunes \citep{barstow2017,pinhas2018_h2o,wakeford2017}. Similarly, emission spectra in the near-infrared with HST WFC3 have led to constraints on H$_2$O abundances in the dayside atmospheres of several hot Jupiters \citep{kreidberg2014b,line2016}. In addition to this, ground-based spectra of directly imaged planets have also provided initial constraints on H$_2$O abundances \citep{todorov2016,lavie2017}. 

The  estimated H$_2$O abundances in transiting and directly-imaged planets are shown in Fig.~\ref{fig:h2o_mass}. Most of the H$_2$O abundance estimates are derived from transmission spectra obtained using HST STIS and WFC3 instruments spanning the optical to near-infrared instruments. While the infrared WFC3 spectral range contains the H$_2$O feature the optical range is necessary to resolve degeneracies with clouds/hazes. Overall, the abundance estimates from transmission spectra across all transiting hot Jupiters known to date are consistent with sub-solar H$_2$O abundances. Whether such low abundances are due to low metallicities in the atmosphere or high C/O ratios is currently unknown. A detailed discussion on these aspects can be found in \citep{madhusudhan2018}. The handful of constraints obtained from emission spectra of a few transiting exoplanets are consistent with both sub-solar and super-solar H$_2$O abundances given their larger uncertainties. On the other hand, directly imaged planets with their superior spectral quality have led to much more precise abundance estimates and indicate super-solar H$_2$O abundances \citep{todorov2016,lavie2017}.

\subsection{Clouds/Hazes}
Inferences of clouds/hazes in exoplanetary atmospheres have been made using varied techniques and instruments. The effect of clouds/hazes on exoplanetary transmission spectra is evident through (a) subdued spectral features of prominent chemical species \citep{deming2013}, and (b) slopes in optical spectra that are deviant from gaseous Rayleigh scattering \citep{pont2013}. In addition, the effects of clouds have also been inferred through optical phase curves of transiting exoplanets \citep{demory2013, shporer2015, munoz2015, parmentier2016} as well as a few reflection spectra \citep{evans2013, martins2015}. At the same time, clouds have also been inferred in directly imaged planets through the modulation of their spectral features in the infrared \citep{marley2015}. A survey of observational inferences of clouds/hazes until a few years ago can be found in \citep{madhusudhan2016}. Here, we outline some key trends.

One of the most surprising findings from the large ensemble of exoplanetary transmission spectra observed is the consistently low spectral amplitudes of H$_2$O absorption features. In the tens of  transmission spectra observed to date, for exoplanets over a wide range of masses and temperatures from cool super-Earths to ultra hot Jupiters, every single one has a H$_2$O feature that is below two scale heights \citep{fu2017,stevenson2016,crossfield2017}. 
This is in stark contrast to expectations, for which a saturated spectral feature in a transmission spectrum is expected to have an amplitude of $\sim$5-10 scale heights. The low spectral amplitudes may indicate either lower H$_2$O abundances than assumed in equilibrium models \citep{madhu2014_3h2o}, the presence of high-altitude clouds obscuring part of the atmosphere \citep{fortney2005,deming2013}, or a high mean molecular weight \citep{line_parmentier_2016}. One way to break the degeneracy is through observations in the optical which, as discussed above, can constrain the scattering mechanisms and hence the presence of clouds/hazes. 

A recent survey of transmission spectra \citep{sing2016} spanning the optical and infrared range provided important constraints on both the properties of clouds/hazes as well as the H$_2$O abundances in ten hot Jupiters. Initial inferences of the data using forward equilibrium models reported the possibility of clouds with no evidence for H$_2$O depletion \citep{sing2016}. However, as discussed in previous section, subsequent studies using atmospheric retrieval methods showed evidence for depleted H$_2$O abundances in most of the planets in the sample, along with varied levels of clouds/hazes \citep{barstow2017,pinhas2018_h2o}. These studies show a diverse range of cloud properties, including the optical slopes, cloud fractions, and cloud-top pressures. Most importantly, these studies show the critical role of the optical range in transmission spectra for constraining both the clouds/hazes as well as the composition. Indeed, other studies that used only near-infrared HST WFC3 spectra show significantly weaker constraints on the cloud parameters and/or the H$_2$O abundances \citep{tsiaras2018}. On the other hand, the low-temperature super-Earths and Neptunes have mostly shown flat spectra in the WFC3 bandpass \citep{kreidberg2014_gj1214b,knutson2014} with a few exceptions \citep{fraine2015,wakeford2017}, which indicate the predominance of clouds in low-temperature atmospheres. 

The potential pervasiveness of clouds/hazes has motivated various studies to devise empirical metrics to quantify the cloudiness in exoplanetary atmospheres. Some studies suggest a metric based on the amplitude of the H$_2$O feature relative to the continuum in adjacent wavelengths \citep{sing2016,stevenson2016}, or the amplitude of the Na/K line centre relative to their adjacent continua \citep{heng2016}. Using known transmission spectra these studies find an intuitive anti-correlation between irradiation and cloud levels, i.e., the likelihood of clouds in atmospheres of hot Jupiters decreases with increasing equilibrium temperature. 

\subsection{Temperature Structures} \label{sec:thermal_inversions}

Temperature profiles of planetary atmospheres provide important insights into radiative processes and their interplay with chemical and dynamical processes. Measurements of pressure-temperature (P-T) profiles in exoplanetary atmospheres are obtained primarily from thermal emission spectra. An emission spectrum probes the brightness temperature in the atmosphere as a function of the wavelength and, hence, the pressure or altitude corresponding to the photosphere at that wavelength. In principle, transmission spectra of transiting planets also provide some constraints on the temperature at the day-night terminator region of the atmosphere but the constraints are relatively weak. This is because there is almost no information in a transmission spectrum about emission from the planet, and the constraint on temperature is primarily through the atmospheric scale height which governs the amplitude of the absorption feature. Therefore, the most stringent constraints on P-T profiles in exoplanetary atmospheres have been obtained for dayside thermal emission spectra of transiting hot Jupiters. A detailed review of observational inferences of atmospheric P-T profiles in exoplanets and their theoretical implications have been discussed in several recent works \citep{madhusudhan2014a,madhusudhan2016,madhusudhan2018}. Here we focus on the latest developments in this area and future directions. 

Recently, new directions are emerging in our understanding of P-T profiles in exoplanets. Originally, constraints on P-T profiles were obtained using only two or more channels of broadband photometry (e.g., at 3.6 $\mu$m and 4.5 $\mu$m). However, current inferences are based on HST WFC3 spectra in the near-infrared in the 1.1-1.7 $\mu$m band along with Spitzer photometry, making the inferences much more robust. Broadly, three classes of P-T profiles have been robustly measured in hot Jupiters: (1) P-T profiles with no thermal inversions, (2) P-T profiles with thermal inversions, and (3) isothermal profiles. Almost all the hot Jupiters observed to date with equilibrium temperatures below $\sim$2000 K show temperature profiles clearly decreasing outward, i.e., with no thermal inversions. More irradiated hot Jupiters show a greater diversity of P-T profiles. Most of these extremely irradiated planets show P-T profiles consistent with isothermal profiles. A handful of these planets also show P-T profiles with no thermal inversions, i.e. with temperatures decreasing outward. And finally, after over a decade of searches, three ultra-hot Jupiters (T$_{\rm eq}$ $\gtrsim$ 2500 K) have been found with detections of thermal inversions in their dayside atmospheres: WASP-18b \citep{sheppard2017, arcangeli2018}, WASP-121b \citep{evans2017}, and WASP-33b \citep{haynes2015}. In summary, while it is now evident that very high temperatures ($\gtrsim$ 2500 K) are a necessary condition for thermal inversions, it is not a sufficient condition since some ultra-hot Jupiters show no signs of thermal inversions \citep{beatty2017b}. 

Observational constraints are beginning to provide new insights into the conditions responsible for thermal inversions in irradiated hot Jupiters. Given that thermal inversions are being seen only in the most highly irradiated planets, the so-called `ultra-hot Jupiters', there is support to the original prediction of TiO/VO being the candidate UV/visible absorbers \citep{hubeny2003,fortney2008} that are causing the inversions. In particular, observations of both WASP-33b and WASP-121b have indicated emission features of TiO \citep{haynes2015,nugroho2017} and VO \citep{evans2017}, respectively. While WASP-18b has not revealed any spectral signatures of TiO/VO, \citep{arcangeli2018} suggested H$^{-}$ opacity as the putative cause. 
Another aspect of the ultra-hot Jupiters with inversions is that none of the spectra show evidence for strong H$_2$O features but do show evidence for CO features. The under-abundant H$_2$O is evident from the muted H$_2$O feature in the HST WFC3 band at 1.4 $\mu$m whereas the significant CO is evident from the excess emission in the Spitzer 4.5 $\mu$m band. These ultra-hot Jupiters have now become the new testing ground for hypotheses on thermal inversions. 

Two competing hypotheses have been put forth to explain the thermal inversions and the lack of strong H$_2$O features from infrared spectra of ultra hot Jupiters. One possibility is a super-solar C/O ratio ($\gtrsim$1) in these atmospheres which can cause low H$_2$O abundance and high CO abundance \citep{madhusudhan2012,moses2013a,sheppard2017}. An alternate explanation is that thermal dissociation of H$_2$O at these high temperatures is responsible for depleted H$_2$O, while molecules such as CO and TiO are relatively more stable \citep{arcangeli2018,parmentier2018,lothringer2018}. The thermal dissociation is also accompanied by H$^{-}$ production. 
This hypothesis can be tested if H$^{-}$ can be retrieved from observed spectra. Currently, inferences of H$^{-}$ (e.g., in WASP-18b) are based on grids of equilibrium models explored to match the data rather than retrieving statistical constraints on the H$^{-}$ abundance from the data \citep{arcangeli2018}. Furthermore, another ultra-hot Jupiter, Kepler-13Ab, which is hotter than WASP-18b, shows a strong H$_2$O feature in absorption indicating the presence of H$_2$O and lack of thermal inversion. This latter anomaly has been explained as due to the high mass and gravity of Kepler-13Ab which enhance cold-trap processes removing inversion-causing species \citep{beatty2017b}. 

Beyond thermal inversions, temperature profiles also provide important constraints on other atmospheric processes. As discussed above, most of the planets observed to date show temperature profiles with no thermal inversions. For irradiated planets, this implies a lack of strong UV/visible absorbers in their atmospheres. On the other hand, for non-irradiated planets such as those observed with direct imaging, the P-T profiles are naturally expected to be devoid of thermal inversions due to the lack of incoming radiation. This is consistent with observed emergent spectra of directly-imaged planets all of which show only absorption features \citep{barman2015, konopacky2013, macintosh2016, bonnefoy2016} and the retrieved P-T profiles also show profiles with no inversions \citep{lavie2017}. 

\subsection{Atmospheric Dynamics}\label{sec:dynamics}

New observations are leading to detailed constraints on atmospheric circulation patterns in hot Jupiters. Constraints on atmospheric dynamics have been obtained with a variety of observations, such as (1) thermal phase curves \citep{zhang2018}, (2) measurements of wind speeds \citep{snellen2010}, (3) eclipse mapping \citep{dewit12}. While thermal phase curves in individual photometric bandpasses have been obtained for a large number of exoplanets, full phase resolved spectra have also been obtained \citep{stevenson2014, kreidberg2018} allowing for detailed retrievals of composition and vertical temperature profiles as a function of orbital phase, as well as the brightness temperature maps as a function of pressure or depth in the atmosphere. The phase curves also provide constraints on day-night temperature contrasts and the location of the day-side hot spot with respect to the sub-stellar point. On the other hand, high-resolution spectroscopy of key species such as CO and Na are providing direct constraints on the wind speeds in the atmospheres of hot Jupiters \citep{snellen2010,louden2015}. 

Observations of thermal phase curves of exoplanets are commonplace today. Phase curves have been observed with a number of facilities, including Spitzer, HST, and Kepler. Most of the observations have been conducted for hot Jupiters, however recently a thermal phase curve was reported for the transiting super-Earth 55 Cancri e \citep{demory2016}. The ensemble of observations has provided two main insights into atmospheric dynamics and energy transport in irradiated exoplanets \citep[e.g.,][]{kataria2016}. Firstly, observations generally confirm the trend of lower energy circulation efficiencies (i.e., increased day-night temperature contrasts) with increasing irradiation (i.e., equilibrium temperature). Secondly, and consequently, the majority of phase curves of hot Jupiters show the hot spot in the dayside atmosphere shifted downwind away from the sub-stellar point, as predicted by General Circulation Models.  

\begin{figure}[t]
\centering
	\begin{subfigure}[b]{0.495\textwidth}
  		\includegraphics[width=\textwidth]{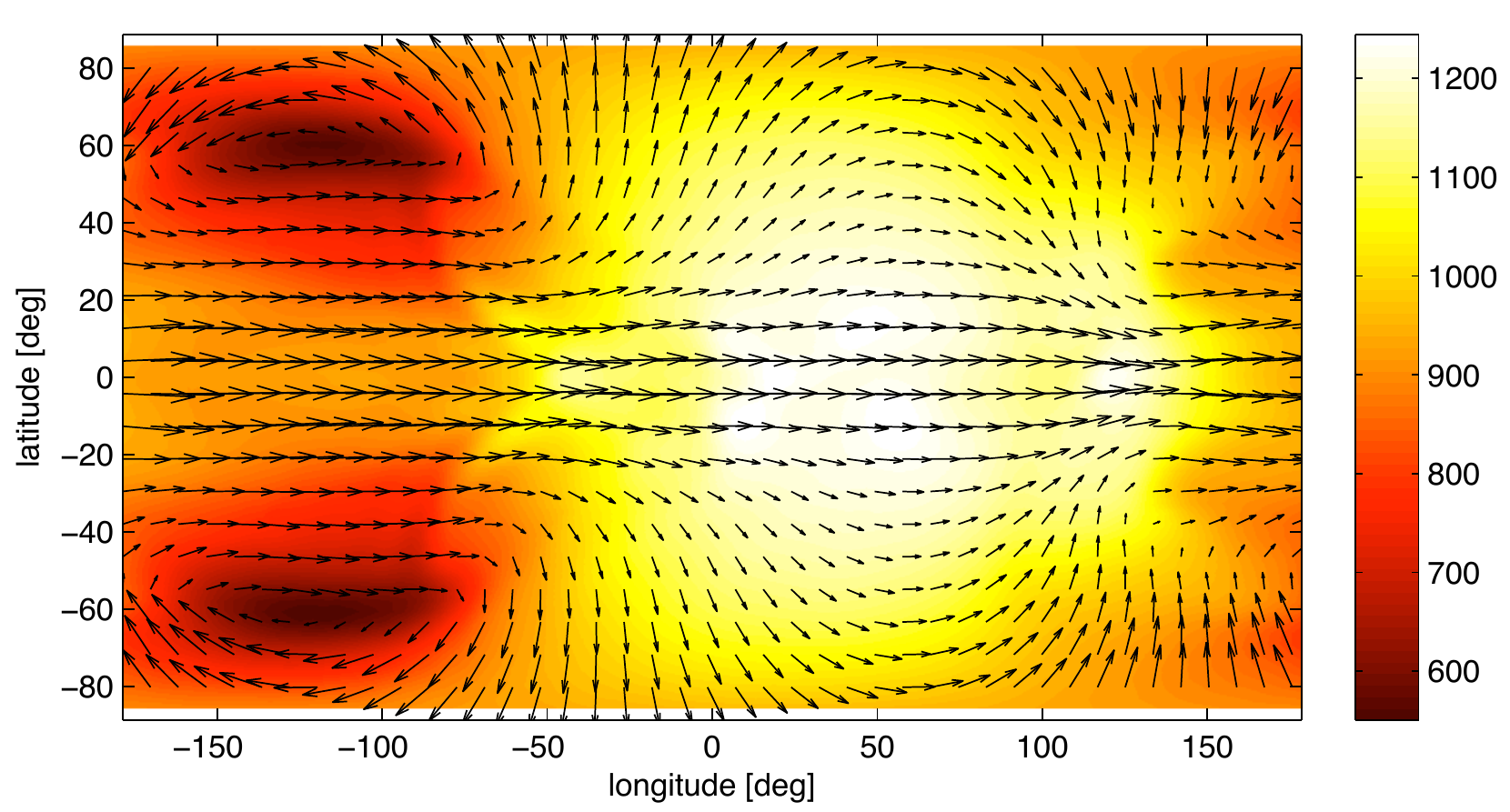}
    \end{subfigure}
	\begin{subfigure}[b]{0.495\textwidth}
  		\includegraphics[width=\textwidth]{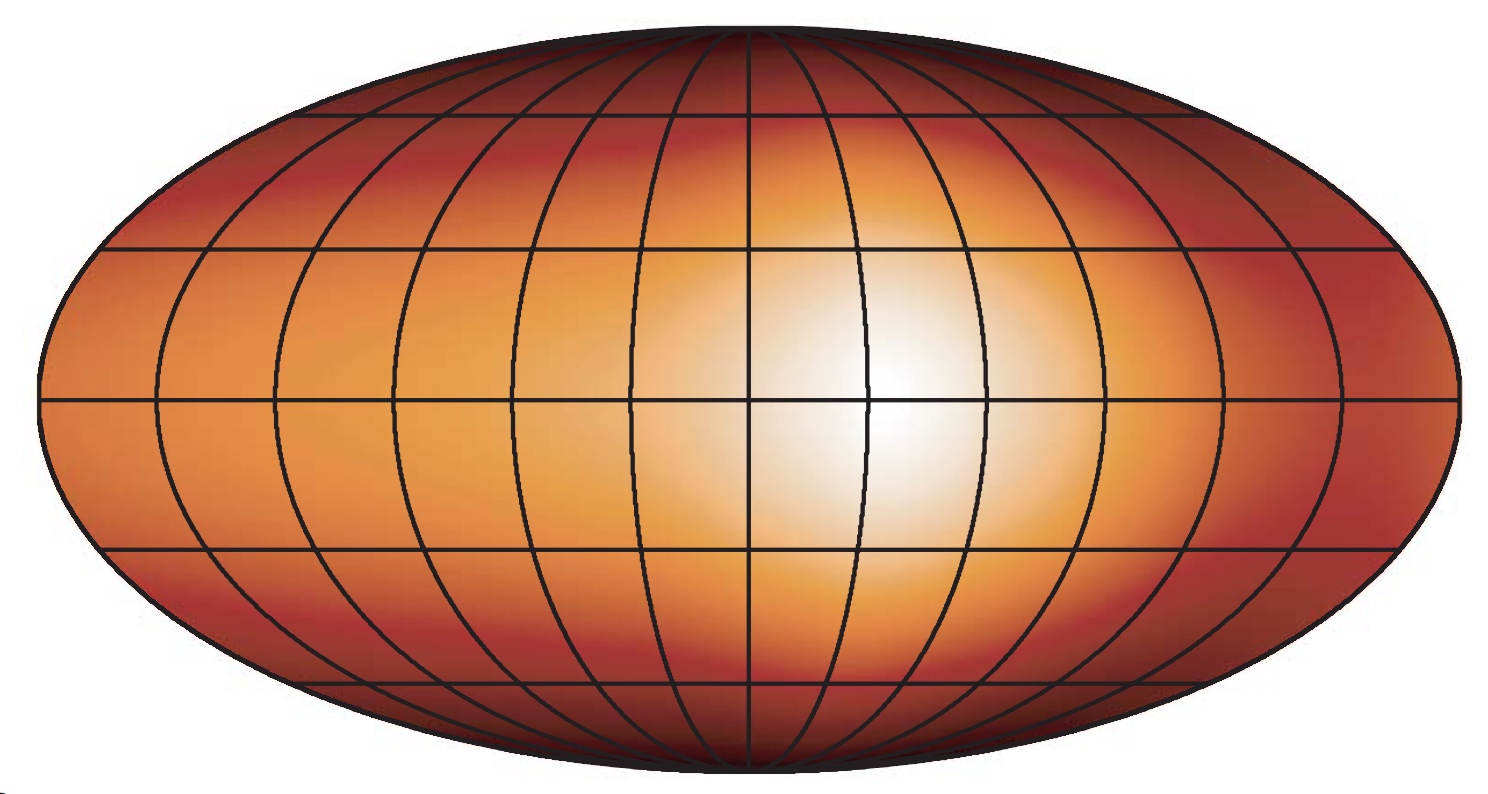}
    \end{subfigure}
\caption{Atmospheric dynamics in hot Jupiters. Left: Theoretical predictions from a GCM of HD 189733b showing an eastward jet and shifting of the hot spot away from the substellar point \citep{showman2011}. Wind vectors are shown with arrows and the coloured temperature scale is in Kelvin. Right: Reconstruction of thermal brightness map for the hot Jupiter HD 189733b using thermal phase curve observations in the Spitzer 8$\mu$m IRAC band \citep{knutson2008}, conforming with the GCM simulations. Figure courtesy Adam Showman.}
\label{fig:dynamics}
\end{figure}

\begin{figure}
\centering
\includegraphics[width=0.6\textwidth]{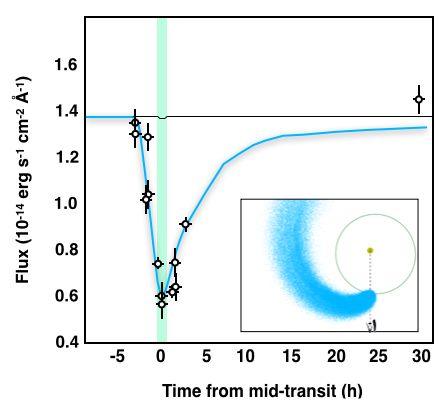}   
\caption{Atmospheric escape of the warm Neptune GJ 436b seen as a strong absorption signature of the stellar HI Lyman-$\alpha$ emission line, based on \cite{ehrenreich2015}. 
The figure shows the Lyman-$\alpha$ transit light curve, where each data point is obtained by integrating the flux from the blue wing of the Lyman-$\alpha$ line as a function of time. The optical transit of GJ 436b is shown by the thin horizontal black line and the in-transit time by the vertical green region. Data are from \cite{ehrenreich2015} and \cite{2017A&A...605L...7L}. The blue curve is the model that best fits the spectra as a function of time, resulting from the particle simulation of \cite{2015A&A...582A..65B}, shown in the inset (where the system is drawn to scale and pole-on). Figure courtesy David Ehrenreich.}
\label{fig:escape}
\end{figure}

High-resolution Doppler spectroscopy of hot Jupiters has led to measurements of wind velocities in their atmospheres consistent with predictions of GCMs. Such measurements were first reported for the hot Jupiter HD 209458b using CO absorption in the near-infrared, observed during transit \citep{snellen2010}, which showed limb-averaged wind speeds of 2 $\pm$ 1 km/s. Similar measurements were made for the hot Jupiter HD 189733b using the Na absorption line observed in a high-resolution transmission spectrum in the optical \citep{wyttenbach2015} and initially reported high wind speeds of 8 $\pm$ 2 km/s. The wind speeds were subsequently revised by \citep{louden2015}, by accounting for the Rossiter-McLaughlin effect, to be 2.3$^{+1.3}_{-1.5}$ km/s and 5.3$^{+1.0}_{-1.4}$ km/s eastward on the leading and trailing limbs of the planet, respectively, suggesting a strong equatorial jet. Such measurements are consistent with predictions from GCMs of hot Jupiters such as HD~209458b and HD~189733b, which show strong eastward equatorial jets and wind velocities of $\sim$1-3 km/s \citep{showman2009,showman2013,zhang2017,lines2018}. Observations such as these provide direct evidence for atmospheric dynamics in hot Jupiters, in addition to thermal phase curves discussed above. 

\subsection{Atmospheric Escape} 
\label{sec:escape}
As discussed in section~\ref{chem_compositions}, transmission spectra of transiting exoplanets in the UV have led to strong detections of several atomic species since the early observations of  \citep[e.g.,][]{vidal-madjar2003}. In particular, detections of exospheric H using observations of Ly-$\alpha$ absorption in the UV have been reported for a number of giant exoplanets, some recent results include \citep{bourrier2013,ehrenreich2015}. Such observations now only allow key insights into the chemical composition of an exosphere but also provide important constraints on hydrodynamic escape processes, the mass loss rates, and the morphology of the escaping cloud. An extensive discussion of this area can be found in the very recent review of \citep{owen2018} and will not be covered in much detail here. We will note, however, that two key directions are emerging in this area. First, it is becoming evident that highly irradiated Neptune-mass planets orbiting low-mass stars provide the best targets for investigating atmospheric escape processes, e.g., the cases of GJ~436b \citep{ehrenreich2015} and GJ 3470b \citep{bourrier2018}. Figure~\ref{fig:escape} shows the example of GJ~436b. Second, the recent detections of He in the near infrared \citep{spake2018} has opened a new avenue to probe exoplanetary exospheres, both from space using HST as well as from ground \citep{allart2018}. These various detections of atomic species have motivated a wide range of theoretical and observational efforts to study escape processes in exoplanets \citep{owen2018}. 

\begin{summary}[SUMMARY POINTS]
\begin{enumerate}
\item The combination of state-of-the-art spectroscopic observations with atmospheric modeling and retrieval techniques has led to a wide range of constraints on atmospheric properties of giant exoplanets.  
\item Chemical detections at high confidence include atomic species (H, He, Na, K, Ti, Fe, C, O, Mg) and molecular species (H$_2$O, CO, HCN, TiO, VO, AlO, CH$_4$).
\item Reliable abundance constraints are available primarily for H$_2$O in transiting giant exoplanets. Transmission spectra of most hot Jupiters   observed in the visible and near-infrared reveal sub-solar H$_2$O abundances. 
\item Clouds/hazes have been inferred in a number of giant exoplanets, primarily from subdued spectral features and non-Rayleigh slopes in optical transmission spectra. 
\item Thermal emission spectra provide key constraints on temperature profiles of giant exoplanets. Thermal inversions have been detected robustly in three ultra-hot Jupiters, with the remaining planets showing no thermal inversions or isothermal profiles.    
\item Atmospheric circulation patterns have been constrained using thermal phase curves, phase resolved spectra, measurements of wind speeds and/or eclipse mapping.
\item Atmospheric escape has been observed in several giant exoplanets with hot Neptunes emerging as optimal candidates for detecing  exospheres. 
\item Overall, atmospheric characterisation is a revolutionary development in the exoplanetary field that only promises to grow as theoretical and observational developments continue. 
\end{enumerate}
\end{summary}

\section{Implications for Planetary Formation} 
\label{sec:planet_formation}

Chemical compositions of planetary atmospheres can provide important insights into their formation and evolutionary mechanisms. As chemical abundances are being measured for exoplanetary atmospheres, an increasing number of studies have been investigating the possibility of using such abundances to answer various open questions about exoplanetary formation. Such studies have mostly focused on giant exoplanets for two main reasons. Firstly, accurate abundance estimates are still only feasibly for giant exoplanets given their higher S/N spetra. Secondly, the H$_2$/He dominated primary atmospheres of giant exoplanets make them important tracers of their accretion history.

\subsection{The Basic Picture}

The ultimate goal of this area is to constrain the primordial formation pathways of exoplanets using their present-day observable chemical abundances. Planetary formation involves a large number of highly complex and stochastic processes. At the outset, therefore, it is imperative to establish which particular aspects of planetary formation can be constrained by atmospheric abundances. The basic set-up of the problem is as follows. The primary assumption is that the initial elemental composition of the protoplanetary disk in which a planet formed is the same as that of the host star, since both are expected to have collapsed from the same protostellar cloud. After the initial collapse, the disk cools in time during which the thermodynamic properties of the disk midplane evolve accordingly. As the disk cools, the snowlines of the various volatile species move inwards towards to the star. As a result, the chemical compositions of both the gas and solids in the disk evolve as a function of time and location in the disk. For example, H$_2$O remains in solids outside the H$_2$O snow line whereas inwards of the snow line it contributes to the gas composition. The same happens with all the prominent volatile species such as CO$_2$, CO, N$_2$ and CH$_4$, which strongly affect the elemental abundance ratios of key elements (e.g., O, C, N) in the disk mid-plane. 

Therefore, the chemical composition of a planet depends on the location and time of its formation in the disk as well as the relative amounts of gas and solids it accretes during its formation. Since the planet may migrate through the disk during formation, the net composition is governed by the cumulative accretion history of the planet over its migration pathway. Beyond this basic picture, there are various other complications. For example, the disk itself can have structural inhomogeneities in the form of gaps and overdensities. The solids may be present in a wide distribution of sizes, from micron-sized dust grains to large planetesimals. The disk composition can be affected by various thermal and photochemical processes.
Moreover, all these processes happen simultaneously, i.e., the planet forms and grows by accreting while migrating in an evolving disk. At the end of this process, all the material accreted by the planet is reprocessed in the planet post formation, finally resulting in the chemical composition observed in its atmosphere today. 

Therefore, the goal of constraining planetary formation processes from atmospheric compositions is a daunting ambition. It is unrealistic to expect for all the involved processes to be constrained solely based on end products. A more reasonable approach is to explore this landscape to assess if any broad regions of the phase space can be constrained or ruled out and to present testable hypotheses given possible atmospheric observations. It is with this spirit that various studies have embarked on this formidable journey. 

\subsection{Compositions of Accreted Material}
The composition of a planet depends on the composition of the material it accretes from the disk both in gas and solid phase. While the gas phase of the disk is mostly composed of H and He with trace quantities of heavy elements, the opposite is true for solids. Therefore, a relatively smaller mass of solids accreted compared to gas can make a disproportionately large contribution to the metallicity of the planet. For example, all the heavy elements in a given amount of solar composition gas amount to only $\sim$1\% by mass. Therefore, the relative amounts of gas to solids accreted by a planet determines the heavy metal content of the atmosphere. On the other hand, the compositions of the gas and dust both evolve in time and orbital distance in the protoplanetary disk \citep{eistrup2018}. While various factors govern the evolution of different disk properties, the most relevant property influencing the composition in the disk mid-plane is temperature which is cooler at larger distances and later times. The mid-plane temperature governs the locations of the snow lines of prominent chemical species which in turn govern whether a particular species is in gas phase or solid phase at a given location. 

\begin{figure}[t]
\includegraphics[width=\textwidth]{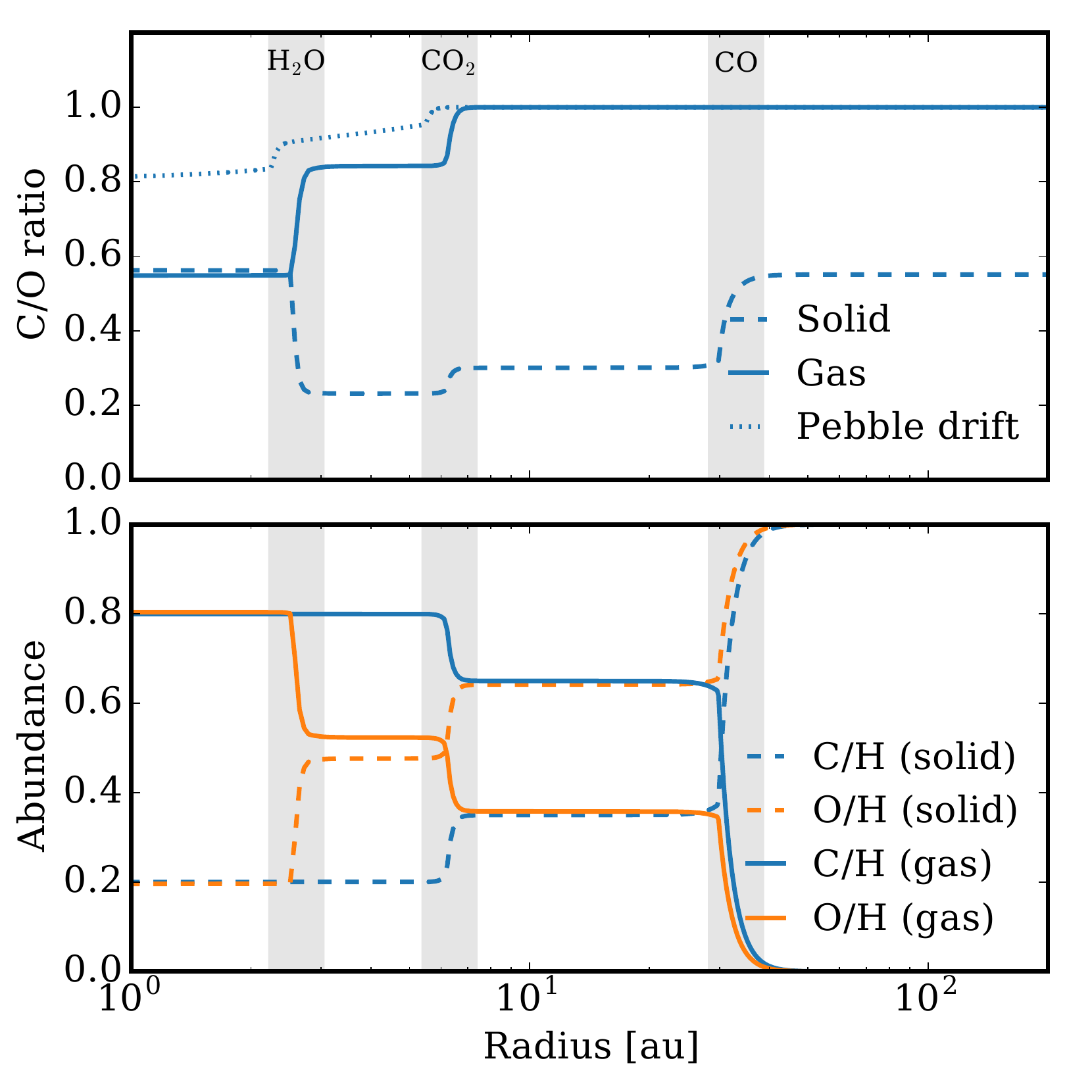}
\caption{Top:Variation of the C/O ratio of the gas and solids in a disc due to freeze-out. The CO, CO$_2$ and H$_2$O snow lines are shown. The width of the snow lines is set by the balance of adsorbtion vs thermal desorption \citep{booth2017}. The dotted line shows a typical case where pebble drift has enhanced the C/O inner the inner region \citep{ali-dib2014, booth2017}. Conversely, the conversion of CO to CO$_2$ and complex organic molecules reduces the C/O ratio of the gas and increases the C/O ratio of the ices (e.g. \citep{eistrup2016, eistrup2018}). Bottom: Abundance of carbon and oxygen relative to solar in solid and gas phases. }
\label{fig:C_to_O}
\end{figure}

Several studies have investigated the compositions of gas and solids that could be accreted by giant planetary atmospheres. The simplest picture is one of a steady state disk where the mid-plane compositions of the gas and solids are determined solely based on the location in the disk relative to the relevant snow lines. Initial studies considered fixed composition disks where the abundances of prominent species (e.g. H$_2$O, CO, CO$_2$) were adopted based on observations of protoplanetary environments and the interstellar medium, or on theoretical models \citep{oberg2011, mousis2011, madhu2014_migration}. With the abundances fixed, the disk temperature at a given orbital distance governs whether each of these species is in gas phase or in solid phase. The sublimation temperatures of these species are such that the H$_2$O snow line is closest to the star (nominally around 5 AU) followed by snow lines of CO$_2$, CO, and other gases such as CH$_4$, N$_2$, and noble gases, in that order. For example, between the H$_2$O and CO$_2$ snow lines, H$_2$O is in solid ice phase whereas all the other volatiles are in gas phase. Therefore, how much gas versus ice is accreted onto the planet at a given location decides how much of each species is accreted onto the planet. The sum-total of all the accreted species contributes to the net elemental abundances of oxygen, carbon, nitrogen, etc. Already from this simple picture it is clear that the oxygen abundance in the gas decreases outward in the disk, with a decrement at each snow line, whereas that in the solids increases. This implies that the C/O ratio in the gas increases whereas that in the solids decreases as a function of distance in the disk \citep{oberg2011}. Figure~\ref{fig:C_to_O} shows the variation of the C/O ratio in the gas and solids in the disk midplane. Therefore, depending on how much gas versus solids a planet accretes at a given formation location it can accrete a wide range of C/O ratios, spanning sub-solar or super-solar C/O ratios; the solar C/O ratio is 0.54 \citep{asplund2009}. Beyond the simple picture above, a host of other processes can influence the gas and solid composition in the disk midplane over time \citep{eistrup2018}. 

\subsection{End-to-end Studies} 
\begin{figure}[t]
\includegraphics[width=\textwidth]{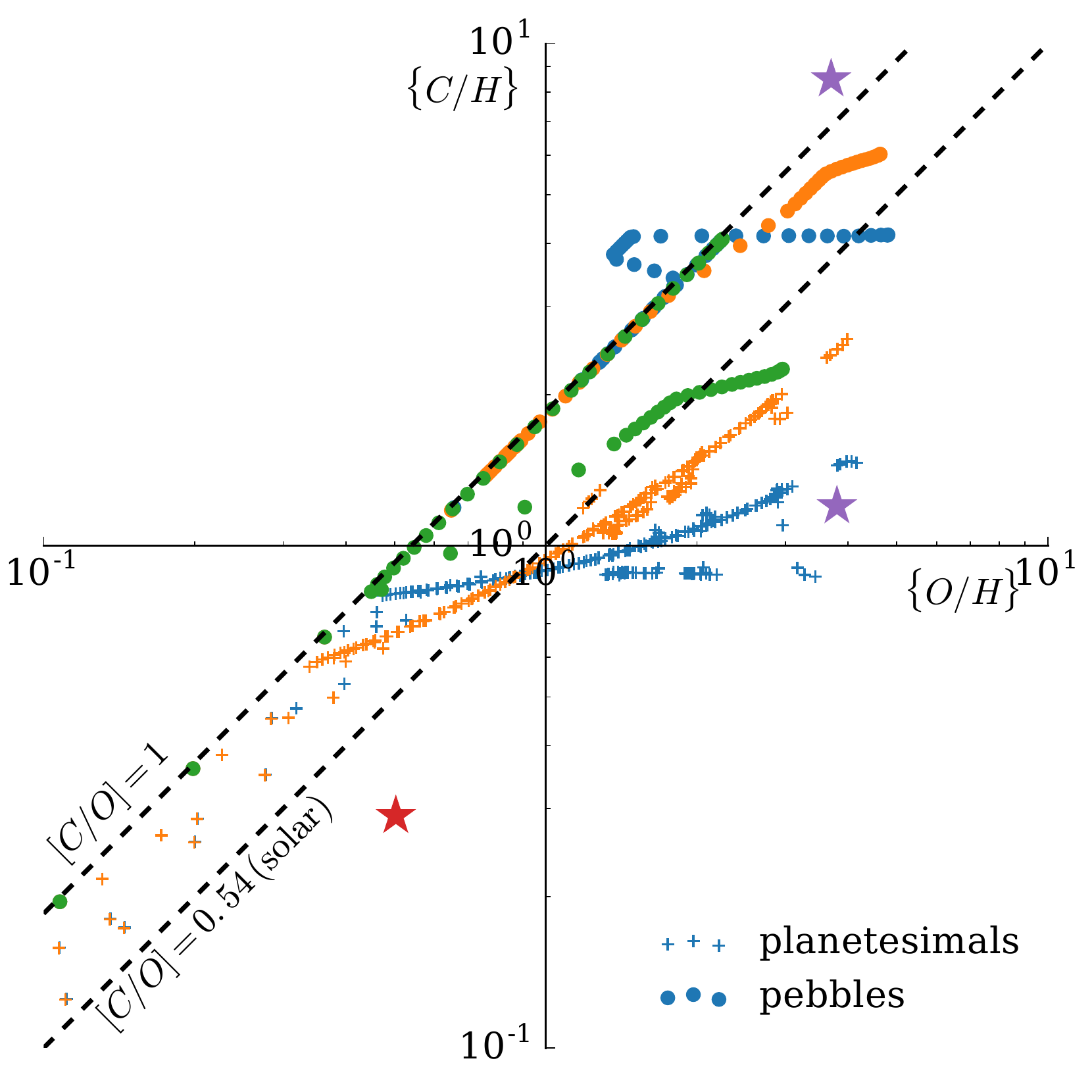}
\caption{Carbon and oxygen abundances arising from different core accretion models of giant planet formation. The circles denote planets formed by pebble accretion, taking into account the chemical evolution driven by pebble migration \citep{booth2017}. Models based on the planetesimal accretion scenario taking into account migration are shown as plus signs \citep{madhu2014_migration}. The point colours denote the chemical abundance models used from \cite{madhu2014_migration}: the case 1 and 2 equilibrium models are shown in blue and orange, and the case 2 no-reactions model in green. The purple stars show the models of Jupiter mass planets from  \cite{mordasini2016}, while the models of \cite{cridland2017} are given by the red star.}
\label{fig:OH_CH}
\end{figure}

In recent years various studies have explored the effects of different formation pathways on the final compositions of giant exoplanets. Detailed reviews on the evolution of this area until recently can be found in \citep{madhusudhan2016,pudritz2018,lammer2018}. Figure~\ref{fig:OH_CH} shows the carbon-oxygen plane with predictions from models assuming different formation pathways. 

Here, after briefly summarising initial works, we focus on the latest developments and the future landscape of this area. The first studies in this direction in the context of giant exoplanets were motivated by the possibilities of measuring elemental abundances and C/O ratios in their atmospheres. Traditionally, solar-system based formation models were used to explain such elemental abundance ratios in giant exoplanets. These argued for local inhomogeneities in formation conditions to explain non-solar abundances \citep{mousis2009,madhusudhan2011}. However, \citep{oberg2011} noted that even in a solar-composition disk, with a C/O ratio of 0.54 \citep{asplund2009}, the C/O ratio of gas and solids in the disk mid-plane changed as a function of radial distance relative to the snow lines of prominent species such as H$_2$O, CO, and CO$_2$. They argued that beyond the CO$_2$ snow line, the gas composition is dominated by CO, giving it a C/O ratio of 1, such that giant planets forming in such regions with predominant gas accretion can have C/O ratios of 1. This study was followed by \citep{madhu2014_migration} who showed that a wide range of C/O ratios and metallicities are possible in giant exoplanets depending on their formation and migration pathways relative to the snowlines. In particular, high C/O ratios ($\sim$1) and sub-stellar metallicities in giant exoplanets were shown to be possible only via mechanisms that did not involve significant solid accretion into the envelope, e.g., disk-free migration or through pebble accretion without core erosion \citep{madhusudhan2017}. Meanwhile, a wide range of C/O ratios and super-solar metallicities were possible for disk migration through abundant accretion of planetesimals \citep{madhu2014_migration, mordasini2016}.

The early studies have been followed by various end-to-end studies which investigated the effects of a wide range of formation conditions on atmospheric compositions of giant exoplanets \citep{mordasini2016,madhusudhan2017,cridland2016,ali-dib2017,booth2017,alibert2017}. These studies have focused largely on hot Jupiters, which are most amenable to atmospheric abundance measurements . These models attempt to accurately capture the formation of the planet, via different mechanisms, while accounting for the chemical inventory of the accreted material. The important differences between the models are in the treatment of the disk chemistry, the specific accretion efficiency of solids versus gas, and in the nature of the accreted material. For example, \citep{mordasini2016} reported an integrated chain model which comprises the formation of the planet by core accretion, migration through the disk, and chemical enrichment caused by planetesimal accretion. They explore two formation pathways, depending on the formation location of the planet relative to the H$_2$O snow line. In this model, the atmospheric elemental composition of the planet is dominated by planetesimal accretion and, regardless of the two formation locations, the composition is predominantly oxygen-rich; owing to significant accretion of H$_2$O ice. On another front, \citep{cridland2016} reported models where the chemistry is treated with an astrochemical model coupled with a disk evolution model. The planet formation is governed by core accretion with gas and planetesimal accretion and formed in specific regions (`traps') in the disk. They also find planets with oxygen-rich C/O ratios of 0.23, owing to an assumed lower C/O ratio in the disc and the assumption of no planetesimal accretion. \citep{venturini2016} also found giant planets with high water abundances, but did not compute C/O ratios. Overall, across these various studies, planets formed via core accretion involving significant planetesimal accretion and migration through the disk result in oxygen-rich composition of the planetary envelopes \citep{madhu2014_migration,mordasini2016,cridland2016}. 

More recent studies have investigated alternate mechanisms that can influence the heavy element content in giant exoplanetary atmospheres. One new direction explored in this context is the formation of giant planets via pebble accretion and migration through the evolving disk with a given chemical prescription \citep{madhusudhan2017}. The important aspect of this formation mechanism \citep{lambrechts2012} is that the solid accretion occurs  predominantly during the formation of the core until the pebble isolation mass is reached, following which gas accretion dominates with almost no solid accretion. In this scenario, the envelope composition is dominated by the gas composition, which is typically metal poor and the accreted solids are sequestered in the core. This leads to generally sub-stellar metallicities and the C/O ratio of the envelope depends on the location of accreted gas relative to the snow lines. In principle, super-stellar metallicities and low C/O ratios can also be attained if the cores are allowed to erode and contribute to the envelope composition \citep{ali-dib2017,madhusudhan2017}. Beyond this picture, however, further studies have considered the effect of pebble drift on the composition of gas in the disk \citep{piso2015,booth2017}. Inward drifting pebbles sublimate at each snow line crossing, thereby enriching the metallicity of the gas inward of the snow line. A planet formed by accreting such metal-rich gas can naturally possess both high metallicities and a wide range of C/O ratios depending on the specific formation pathway. Overall, the pebble accretion paradigm provides a natural way to explain a wide range of metallicities as well as C/O ratios in giant exoplanetary atmospheres. 
A new generation of studies are now beginning to investigate the effect of complex disk chemistry on the compositions of the planets formed via different mechanisms. In previous studies, the treatment of chemistry in the disk was simplistic based on fixed condensation fronts depending on the mid plane temperature \citep{oberg2011,madhu2014_migration}. New studies considering full chemical kinetics along with the evolving disk show significant evolution in the gas and ice compositions (e.g. C/O ratios) of the disk mid-plane beyond what fixed chemical prescriptions assume \citep{eistrup2018}. Future end-to-end models including such full chemical treatment may find a wider diversity of planetary compositions than currently predicted. A recent study with an end-to-end model combined a full chemical model with a hydrodynamical model of planet formation by gravitational instability \citep{ilee2017}. They suggested the possibility of dust grains, along with their volatiles, sedimenting to the cores and preventing the enrichment of their envelopes with the volatiles. Therefore, planetary envelopes formed via gravitational instability may not entirely reflect the bulk composition of the formation locations. Finally, new studies are also suggesting that besides the carbon and oxygen abundances, the nitrogen abundance may also be an important diagnostic of planetary formation pathways \citep{piso2016,booth2018}.

\begin{summary}[SUMMARY POINTS]
\begin{enumerate}
\item The key goal is to use elemental abudances in exoplanetary atmospheres to constraint their formation and migration history. Basic premise: elemental composition of protoplanetary  disk same as the protostellar nebula. 
\item Composition of the disk (e.g. C/H, O/H, C/O ratios) in gas and solids change as a function of orbital distance and age, with snow lines of H$_2$O, CO, and CO$_2$ governing key transitions in elemental abundances. Further in the disk beyond the CO$_2$ snowline the C/O ratio in the gas approaches 1. 
\item Giant planet composition depends on the relative amounts of gas vs solids accreted by the planet from different locations in the disk relative to the snow lines. 
\item Giant planet formation via core accretion with significant planetesimal accretion and migration through the disk leads to super-solar metallicities and sub-solar (oxygen-rich) C/O ratios, i.e., C/O $\lesssim$ 0.5.
\item Giant planet formation beyond the CO/CO$_2$ snow lines but with disk-free migration can lead to sub-solar metallicities and carbon-rich compositions with C/O approaching 1. 
\item{Giant planet formation by pebble accretion can lead to a diverse range of metallicities and C/O ratios depending on the efficiency of core erosion as well the contribution of pebble drift to the disk composition.}
\end{enumerate}
\end{summary}

\section{Habitable Planets and Biosignatures}
\label{sec:habitability}
The detection of a biosignature in the atmosphere of a terrestrial-size planet represents 
the holy grail of exoplanetary science. The current population of discovered transiting exoplanets contain 
dozens of rocky exoplanets in their habitable zones \citep{HZ_planets}. It is conceivable that a chemical signature in the atmosphere of a habitable exoplanet may be discovered within the next decade with large facilities in space (e.g. JWST) and ground (e.g. E-ELT). However, whether or not the detected chemical will be a robust signature of life is 
a subject of much debate \citep{meadows2018a,catling2018}. Nevertheless, the prospects of 
finding a potential biosignature in the atmosphere of a terrestrial exoplanet are promising 
given the rapid increase in new discoveries of such planets and the new observational facilities 
on the horizon. Recent reviews of various aspects of this area can be found in \citep{meadows2018a,catling2018,fujii2018,seager2018, madhusudhan2016}

\subsection{Habitable Planets}

One of the most revolutionary findings from exoplanet detection surveys has been the high occurance rate of rocky exoplanets in the solar neighbourhood \citep{fulton2017}. Earth-size planets are now known to be very common, and more so around low-mass stars. In particular, dozens of terrestrial-size exoplanets are known with equilibrium temperatures conducive for liquid water to exist on the surface. In principle, these planets may be be classed as habitable-zone planets, though the extent of a habitable zone is subject to the planetary interior and atmospheric properties as well as the astrophysical conditions assumed. A detailed review of the various factors affecting habitability of terrestrial exoplanets can be found in various recent works \citep{gudel2014,seager2018,kopparapu2017,grenfell2007,kaltenegger2017,selsis2008,haghighipour2007}. These factors include the atmospheric and geophysical conditions in the planet, its orbital parameters and evolution, the nature and evolution of the host star and its environment, magnetospheric protection, and the planet's formation history, among other factors. The dozens of potentially habitable planets now known are found orbiting mostly late-type K and M stars, meaning their environmental conditions may be expected to be very different and diverse compared to the terrestrial experience. An exact Earth analog is yet to be discovered. 
From an observational standpoint, however, planets orbiting late-type stars provide a fortuitous opportunity to characterise their atmospheres. Their small stellar sizes provide large planet-star contrasts, both in radius and flux, making them conducive for transit spectroscopy. This `small star opportunity' has emerged to be the cornerstone in the search for habitable planets and biosignatures in the near future. In particular, the recent discoveries of habitable-zone planets orbiting nearby stars such as TRAPPIST-1 \citep{gillon2016} and Proxima Cen \citep{anglada-escude2016} provide new impetus in this direction. As discussed below, discovering biosignatures in the atmospheres of such planets may be within the reach of upcoming observational facilities.

\subsection{Biosignatures}

The ultimate breakthrough in exoplanetary science will be the detection of a biosignature in the 
atmosphere of a rocky habitable-zone exoplanet. But, what is a biosignature? Nominally, an ideal 
biosignature gas would need to satisfy some natural conditions \citep{seager2016}, such as (a) it should not have any false positives, i.e., should not be a product of non-biological mechanisms, (b) it should have strong enough spectral features to be detectable, and (c) it should be abundant enough to be detectable.
Traditionally, the prominent biosignatures based on the Earth's atmosphere were thought to be 
O$_2$, O$_3$, N$_2$O, and CH$_4$; though CH$_4$ is known to be produced in minor quantities 
geologically \citep{catling2018}. By far, the most promising biosignature for Earth-like planets
had originally been considered to be O$_2$ and, hence, O$_3$. However, recent studies suggest the 
possibility of abiotic mechanisms which can also produce O$_2$ in detectable quantities \citep{meadows2018a}. 
It may be safe to say that currently there is no single molecule that may qualify as a unique 
biosignature upon detection in extraterrestrial atmospheres. Nevertheless, recent studies are suggesting 
several approaches for progress in this direction. Firstly, any assessment of a biosignature has to 
take into account various aspects of the environment that are required for the planet to be conducive to life \citep{catling2018}. These include characterisation of the stellar host and bulk properties of the 
planet as well as the atmospheric `climatic' conditions. Ultimately, a probabilistic measure of life may 
be more plausible than a binary inference \cite{catling2018}. Furthermore, while there is no single ideal molecule, 
the combination of multiple species (e.g. O$_2$ and CH$_4$) may be a potential biosignature under the 
given conditions. In this regard, a detection of O$_2$ and CH$_4$ and/or N$_2$O along with liquid H$_2$O 
on a habitable-zone planet, i.e. an almost exact Earth analogue, may be a sure sign of life. 

While designing metrics centered on terrestrial experience may be a convenient starting point, it will 
likely serve us better to be open to surprises. As is common experience in exoplanetary science, reality has 
rarely conformed to our expectations. From planetary detections to atmospheric studies, most of the findings 
to date have defied expectations. Initial studies show that the possible space of biochemical byproducts may 
comprise of numerous chemical species \citep{schwieterman2018,seager2018}, and may indeed be non-denumerably 
infinite. For example, species such as O2, O3, N$_2$O, CH$_4$, CO$_2$, CH$_3$SH, CH$_3$Cl, C$_2$H$_6$ and NH$_3$, may all be potential bio-signatures in Earth-like conditions \citep{schwieterman2018,seager2018}. 
In addition, different stellar environments may reduce or increase the chances of potential biosignatures. 
For example, while planets around M dwarfs may be subject to an extremely harsh UV environment which can be 
catastrophic for terrestrial-like life \cite{segura2010,kaltenegger2017}, the same environments may also 
be conducive for abiogenesis, producing species essential for primordial life such as HCN \citep{rimmer2017,rimmer2018}. 

\subsection{Observational Prospects} 

A prudent approach to searching for biosignatures in terrestrial exoplanets is likely to be one that is driven by observational capability rather than terracentric predictions of an ideal 
biosignature. Several recent studies have discussed the observational capabilities of upcoming facilities that would be sensitive to detecting chemical signatures in rocky exoplanets \citep{kaltenegger2017,seager2013, schwieterman2018, fujii2018}. 
Atmospheric characterisation of Earth-like planets around sun-like stars may be beyond the capabilities of current and upcoming facilities. However, a more promising pathway that has emerged in recent years is the atmospheric characterisation of planets orbiting low-mass stars such as TRAPPIST-1 
\citep{gillon2016} and Proxima Centauri \citep{anglada-escude2016}. Recent studies show that dedicated 
spectroscopic observations with JWST could take us within the reach of detecting chemical 
signatures for habitable planets in both TRAPPIST-1 \citep{morley2017, batalha2018} and Proxima Cen b \citep{snellen2017}. Furthermore, these atmospheres may also be probed by large ground-based facilities (e.g. E-ELT) using a combination of high-resolution spectroscopy and high-contrast imaging \citep{snellen2015,rodler2014}. New studies are underway for a future generation of facilities such as LUVOIR \citep{bolcar2016} and HabEx \citep{mennesson2016} that would enable characterisation of potential biospheres of Earth-like planets around sun-like stars \citep{wang2018}.

\begin{summary}[SUMMARY POINTS]
\begin{enumerate}
\item The detection of an atmospheric biosignature in a rocky exoplanet represents the holy grail of the field. 
\item Defining a unique biosignature remains a theoretical challenge, but several candidate molecules have been suggested. 
\item Detection of a potential biosignature may be within the reach of upcoming observational facilities. 
\end{enumerate}
\end{summary}

\section{Future Landscape}

It is clear that the future of exoplanetary science lies in detailed characterisation of exoplanetary atmospheres. Numerous surveys both from space and on the ground are geared towards discovering planets orbiting nearby bright stars which enable detailed atmospheric spectroscopy of the planets. Current and upcoming transit surveys include K2, TESS, and CHEOPS from space and ground-based surveys such as SuperWASP, NGTS, KELT, SPECULOOS, and MEARTH, followed by the PLATO mission on a longer term. In parallel, direct-imaging surveys with instruments such as GPI, SPHERE, and SCExAO aim to discover long-period exoplanets around young nearby stars. These surveys promise a large sample of targets for detailed comparative characterisation of exoplanetary atmospheres over the next decade. On the other hand, the prospects of various new observational facilities for atmospheric spectroscopy of exoplanets are equally exciting. In the imminent future, JWST will play a pivotal role in revolutionising the field. In the late 2020s space-based facilities such as ARIEL will enable atmospheric characterisation of large populations of exoplanets, and large ground-based telescopes such as E-ELT, GMT, and TMT, may enable detections of biosignatures in atmospheres of habitable-zone exoplanets. Here we discuss the future landscape of the field, focusing specifically on observational facilities at present and in the imminent future. 

\subsection{Exoplanetary Atmospheres with Current Facilities}
As discussed in this work, current facilities are already capable of providing detailed insights into exoplanetary atmospheres. What can we expect from these facilities in the future? HST, which has been the workhorse of the field in recent years, will continue to be a powerful facility. Prior to JWST, which is currently expected to launch in 2021, HST is the only space-based facility for transit spectroscopy over a broad spectral range, from UV to near-IR. As discussed above, spectra over the broad spectral range from visible to NIR have been obtained with HST for over ten giant exoplanets, albeit with variable data quality. Similarly, spectra are becoming available for tens more transiting exoplanets obtained with one or more HST instruments albeit with generally limited data quality. These kinds of survey programs are beneficial to provide a general census of exoplanetary spectra over a large sample \citep{sing2016,tsiaras2018,fu2017}, but are of limited utility for obtaining stringent constraints on relevant atmospheric properties of any individual planet. As discussed in \citep{madhusudhan2018}, H$_2$O abundances have been measured to better than 0.5 dex precisions in only a handful of exoplanets. Retrievals of high-precision chemical abundances and other atmospheric properties from transit spectra require uncertainties on data that are nearly 20 ppm \citep{deming2013,madhu2014_3h2o,kreidberg2014b,macdonald2017}. While such precisions can be obtained in single transit/eclipse observations for planets orbiting bright host stars such as HD~209458b \citep{deming2013}, fainter host stars require coadding spectra from multiple events, as demonstrated for WASP-43b \citep{stevenson2014,kreidberg2014b}. This is the approach required with HST, i.e., dedicated efforts for high-precision (20 ppm) transit spectroscopy over the entire HST spectral range of a sizeable sample of exoplanets. Such a program has the potential to obtained high precision chemical abundances of H$_2$O, and other possible species (e.g. CH$_4$, HCN, NH$_3$, TiO/VO, Na, K), P-T profiles, constraints on clouds/hazes for a large sample of exoplanets to enable comparative atmospheric characterisation.

At the same time, as discussed earlier, ground-based facilities have started to provide high-precision spectra for both close-in planets as well as directly imaged planets on large orbital separations, as shown in Fig.~\ref{fig:di_panorama}. In particular, ground-based transit spectroscopy at both low-resolution and high-resolution will continue to provide robust detections of several chemical species such as Na, K, TiO, and He \citep[e.g.,][]{sedaghati2017,nikolov2018}. Such detections can be made using instruments on existing large ground-based telescopes (e.g. VLT and GTC) as well as high-resolution spectrographs on medium-size telescopes such as the HARPS spectrograph \citep{hoeijmakers2018}. In addition, high resolution Doppler spectroscopy in thermal emission with large ground-based telescopes promise high confidence detections of chemical species. As discussed earlier, this technique has been instrumental in strong detections of key species such as H$_2$O, CO, TiO, and HCN in several hot Jupiters using the CRIRES instrument on the VLT. The upcoming CRIRES+ spectrograph on VLT along with other high-resolution spectrographs on medium-size (4m class) telescopes promise new opportunities for the future \citep{brogi2017}. Ultimately, the combination of high-resolution Doppler spectroscopy along with low-resolution transit spectroscopy will provide the best constraints possible. Similarly, the combination of the cross-correlation technique  with high-contrast imaging provides new promise for medium-high resolution spectroscopy at extreme flux contrasts\citep{snellen2015,hoeijmakers2018}. These advancements will naturally be further enhanced with the next generation of large ground-based facilities such as the E-ELT, GMT, and TMT. 

\subsection{Exoplanetary Atmospheres with JWST}

The James Webb Space Telescope (JWST) has the potential to revolutionise the study of exoplanetary 
atmospheres. The key advantages of JWST for exoplanet spectroscopy are apparent - its large aperture, 
and hence high sensitivity, and wide spectral range, particularly in the infrared. Several studies have  
investigated the science that can be pursued with JWST \citep{greene2016, deming2009, bean2018}. 
Here, we discuss some key highlights and some strategic factors that need to be considered to maximise 
the potential of JWST for atmospheric characterisation of exoplanets. 

\begin{figure}[h!]
\includegraphics[width=\linewidth]{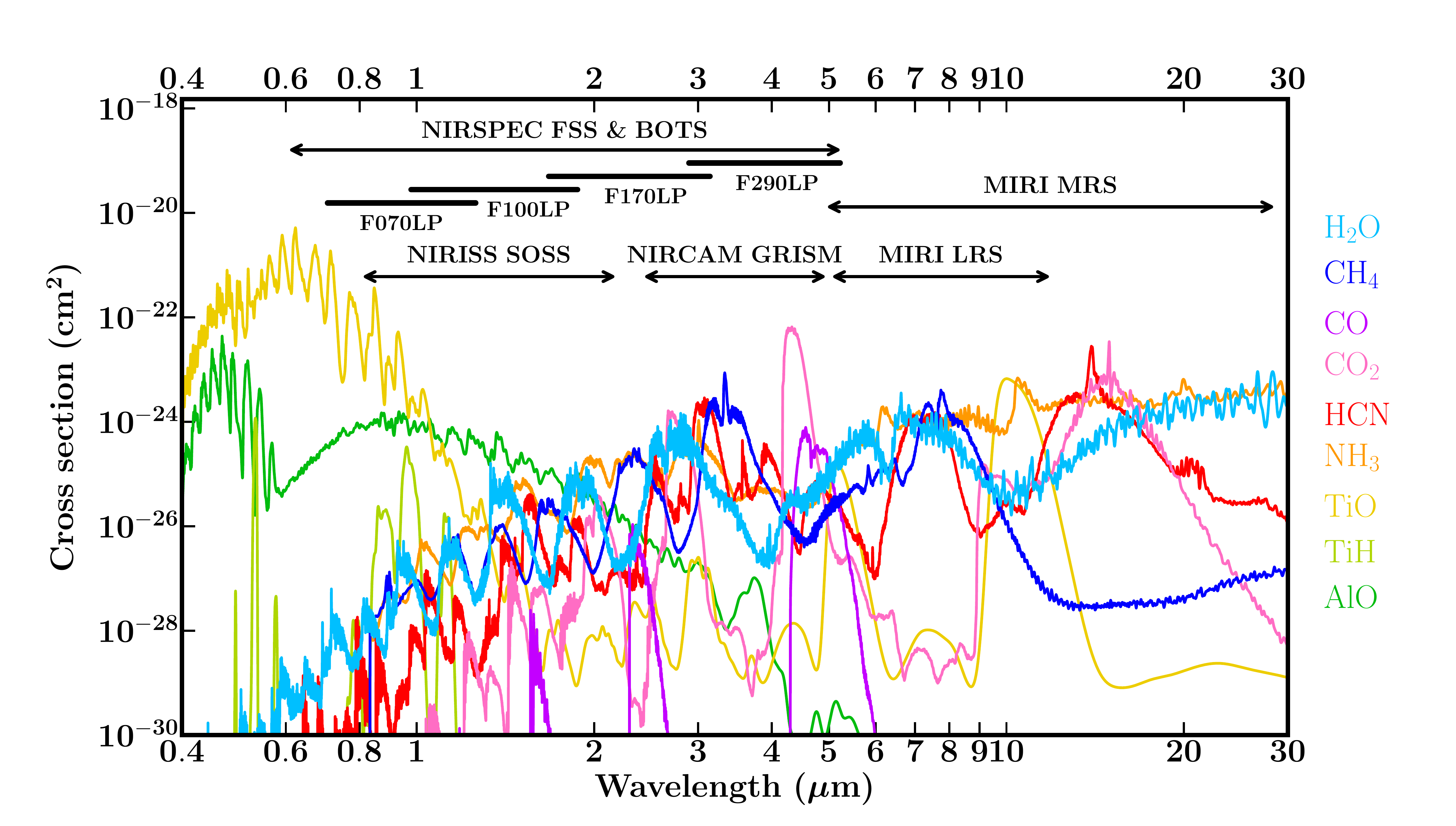}
\caption{Cross-sections of various molecules across optical and infrared wavelengths. Spectral ranges covered by certain modes of JWST's four instruments (NIRSpec, NIRISS, NIRCam, and MIRI) are shown for comparison. For NIRSpec, the wavelength coverage of individual filters is also shown. JWST's extensive spectral coverage will enable detailed chemical characterisations of exoplanetary atmospheres.}
\label{fig:xsections}
\end{figure}

JWST will clearly revolutionise our understanding of giant planets which will be the most conducive planets 
for atmospheric characterisation as already demonstrated with current facilities. This is most apparent when 
considering that the range of species that JWST will be able to observe in hot Jupiters may supersede 
what we know for Jupiter in our own solar system, which is currently the most studied giant planet. 
The wide spectral range of JWST spans the absorption features of a range of chemical species, as shown 
in Fig.~\ref{fig:xsections}. All the prominent molecular species containing the key volatile elements (e.g. O, C, and N) 
such as H$_2$O, CO, CH$_4$, CO$_2$, HCN, C$_2$H$_2$, and NH$_3$, have strong spectral features in the 
JWST spectral range. The mixing ratios of these species can provide unprecedented constraints on the 
corresponding elemental abundance ratios, e.g., C/H, O/H, N/H, C/O, N/O, etc. These molecular and elemental 
ratios have the potential to provide a wide range of constraints on numerous atmospheric processes and 
planetary formation mechanisms, as discussed in this work. Moreover, the spectral range of JWST, as seen in Fig. \ref{fig:xsections},
also contains strong signatures of various molecules containing refractory elements (e.g. TiO, VO, AlO, TiH, etc) 
which provide important constraints on both the metal content in the atmospheres as well as the possibility 
of aerosols comprised of these species, depending on the temperatures. Most of these species are not 
measurable for giant planets in the solar system due to their low temperatures. On the other hand, JWST will 
be able to measure these for planets over a wide mass range, possibly from super-Earths to super-Jupiters. 

\begin{figure}[h]
\includegraphics[width=1.0\linewidth]{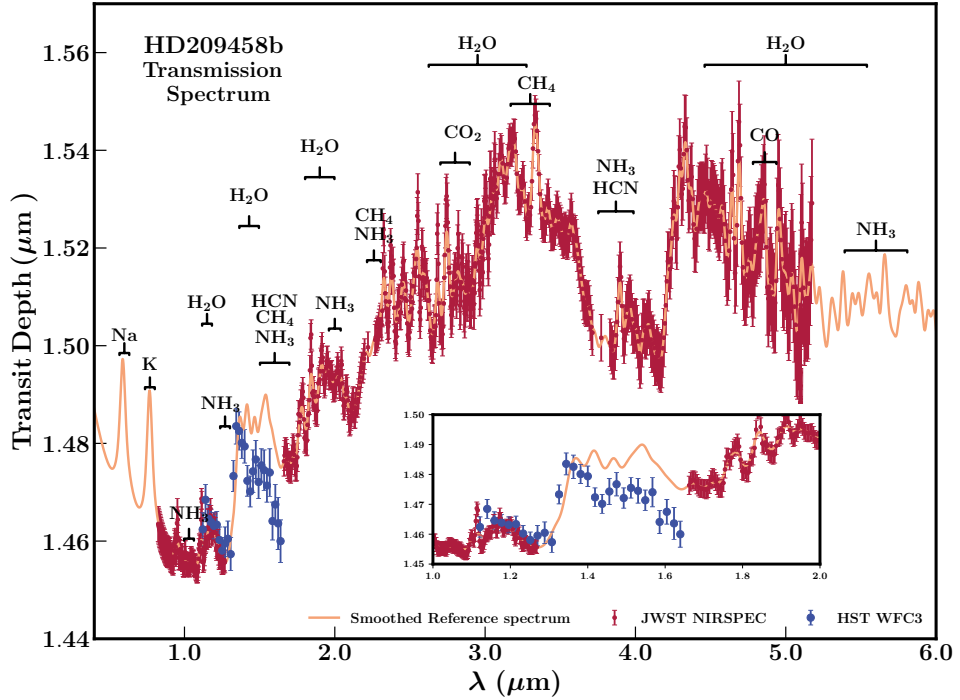}
\caption{Simulated JWST/NIRSpec data based on the model spectrum of a planet resembling HD 209458b . Simulated data is shown by dark red points, while the model spectrum is shown in pink. Data is simulated for the G140H, G235H and G395H high-resolution grisms available with NIRSpec. The native-resolution simulated data points have been binned to a lower resolution for clarity. In the inset, real HST/WFC3 data for HD 209458b (blue points and error bars) is shown for reference. The chemical features probed in this spectral range are labelled with brackets; atmospheric observations with JWST will allow higher resolutions and smaller uncertainties than present capabilities, resulting in more robust and varied chemical detections.}
\label{fig:hd209_simulated}
\end{figure}

While JWST will readily detect various chemical species in exoplanetary atmospheres, precise quantitative 
constraints on the atmospheric properties will require judicious planning. The 
atmospheric properties including the chemical abundances, temperature profiles, and other properties, are 
determined from spectra using atmospheric retrieval methods. It is conceivable that chemical abundances and 
ratios (e.g. H$_2$O, C/O ratio, etc) can be determined to precisions within 0.5 dex \citep{greene2016,batalha2018}, 
more so considering that retrievals on HST spectra are already providing best precisions on H$_2$O abundances 
of $\sim$0.5 dex \citep{madhu2014_3h2o,macdonald2017,wakeford2017}. However, the precision of the determined chemical abundances, depends both on the spectral range as well as the precision, while ensuring that 
the desired chemical species has strong signature in the observed range. Firstly, as shown in Fig.~\ref{fig:hd209_simulated}, the 
spectral ranges of the near-infrared instruments (NIRISS and NIRSPEC) are composed of multiple modes 
each of which covers a narrow range, typically 1-2 $\mu$m wide, and can be observed only one at a time. Consequently, obtaining the full spectral range of JWST would require a significant investment of JWST time on any one target. Therefore, judicious choices need to be made in choosing the observing modes which depend on a range of system properties including the stellar brightness, planetary bulk properties (e.g. size, gravity, temperature), expected atmospheric composition and spectral amplitude. Various studies have investigated the information content in JWST spectra as a function of these properties \citep[e.g.,][]{greene2016,batalha2018,molliere2017,
rocchetto2016} using simulated JWST data. For example, \citep{batalha2018} show that for a given amount of observing time with JWST NIRISS and NIRSpec the information content in an observed spectrum of a certain molecule in multiple modes is higher than that observed in a single mode at high precision. On the other hand, for certain host stars a broader coverage over 1-5 $\mu$m range can be achieved with the NIRSpec in GRISM mode. In principle, the MIRI instrument has significant larger spectral range in each of two models in the 5-28 $\mu$m but have lower resolution and fewer relevant spectral features compared to the near-infrared. 

Another challenge for transit spectroscopy with JWST lies in the possibility of cloudy atmospheres. As discussed in this work, transmission spectra have suggested the presence of clouds/hazes in transiting exoplanets over a range of temperatures and masses, revealed through low-amplitude H$_2$O features as well as non-Rayleigh spectral slopes in the 
optical \citep{sing2016,fu2017,tsiaras2018}. Current retrieval codes are able to retrieve the chemical 
abundances from such spectra in addition to constraining the cloud/haze properties \citep{barstow2017,macdonald2017}. 
However, such joint constraints require transmission spectra in visible wavelengths to resolve the degeneracies 
between the chemical abundances and clouds/hazes. While this has been possible with HST, using the STIS spectrograph, 
the same will be lacking for JWST which does not have significant spectral coverage in the optical. Therefore, 
complementary transmission spectra in the visible observed with HST and/or large ground-based telescopes will 
be required to accurately retrieve chemical compositions of cloudy atmospheres using JWST transmission spectra. 
The recent successes of HST and high-precision ground-based transmission spectroscopy of giant exoplanets \citep{sedaghati2017,nikolov2018,chen2018} provide promising prospects in this direction. On the other hand, for 
many of the irradiated giant exoplanets dayside thermal emission spectra may provide a more viable means to constrain 
their atmospheric compositions, rather than transmission spectra. The dayside spectra are less influenced by clouds owing to the higher irradiation/temperatures as well as the fact that an opaque cloud deck essentially acts as a photosphere and the compositions constrained are those above the cloud deck by default. Moreover, infrared spectra 
of several hot Jupiters with HST and Spitzer, as well as with high-resolution ground-based spectra have already 
revealed clear signatures of molecular features and temperature gradients suggesting the lack of strong interference 
due to clouds/hazes on the dayside spectra. Besides chemical compositions, JWST will provide unprecedented constraints on a host of other atmospheric properties of exoplanets. In addition, JWST will provide the first opportunity to characterise the atmospheres of super-Earths and terrestrial-size exoplanets in detail, especially for planets orbiting low-mass stars \citep{snellen2017}.

\section*{DISCLOSURE STATEMENT}
The authors are not aware of any affiliations, memberships, funding, or financial holdings that might be perceived as affecting the objectivity of this review. 

\section*{ACKNOWLEDGMENTS}
At the outset I would like to thank the countless colleagues in the field who have contributed to this exciting frontier of exoplanetary science. I would like to thank the editors at ARAA for the opportunity to serve the field with this review, and Debra Fischer for her very helpful comments on the review. The volume of literature on this emerging area of exoplanetary atmospheres is both exciting and astounding at the same time, a testament to the exponential progress we are witnessing. As such, given the space and time available for this review, choices had to be made about which aspects to cover at what level. I therefore sincerely apologize to all those colleagues whose work may not have received coverage in this review. I would like to deeply thank my students Luis Welbanks and Anjali Piette for their invaluable contributions to the figures, tables, and sorting the references in this review, Arazi Pinhas for help with the tables, some figures and references, Richard Booth for help with Figs. 8 \& 9, and Siddharth Gandhi for help with some of the references. I would also like to thank Adam Showman and David Ehrenreich for their very helpful comments and contributions to sections \ref{sec:dynamics} and \ref{sec:escape}, respectively.



\end{document}